\documentclass[letterpaper]{article}

\usepackage[T1]{fontenc}

\usepackage{geometry}
\usepackage{setspace}

\usepackage[
    style = chem-acs,
    articletitle=true,
    maxnames=99,
    minnames=99,
    doi=true
]{biblatex}

\usepackage{graphicx}
\usepackage{float}

\usepackage{caption} 
\usepackage{booktabs} 
\usepackage{multirow} 
\usepackage{array} 
\usepackage{tabularx} 
\usepackage{adjustbox} 
\usepackage{siunitx}

\usepackage{amsmath} 
\usepackage{amssymb} 
\usepackage{bm}

\usepackage[version = 4]{mhchem} 

\usepackage[hidelinks]{hyperref}
\usepackage{pdfpages}

\usepackage{authblk}
\author[1,2]{Rafa{\l} Topolnicki*}
\author[3,4]{Jan Navr\'atil}
\author[4,5]{Piotr B{\l}o\'nski*}
\affil[1]{Institute of Experimental Physics, University of Wroc{\l}aw, pl. M. Borna 9, Wroc{\l}aw, 50-204, Poland}
\affil[2]{Dioscuri Center in Topological Data Analysis, Institute of Mathematics, Polish Academy of Sciences, ul. {\'S}niadeckich 8, Warsaw, 00-656, Poland}
\affil[3]{Regional Centre of Advanced Technologies and Materials, Czech Advanced Technology and Research Institute (CATRIN), Palack\'y University Olomouc, \v{S}lechtitel\r{u} 27, 779 00 Olomouc, Czech Republic}
\affil[4]{Department of Physical Chemistry, Faculty of Science, Palack\'y University Olomouc, t\v{r}. 17 listopadu 12, 779 00 Olomouc, Czech Republic}
\affil[5]{IT4Innovations, V\v{S}B--Technical University of Ostrava, 17. listopadu 2172/15, 708 00 Ostrava-Poruba, Czech Republic}

\title{Can end-to-end learning from raw electronic structure explain magnetic anisotropy?}
\date{*Email: rafal.topolnicki@uwr.edu.pl; piotr.blonski@upol.cz}

\begin{document}


\maketitle
\begin{abstract}
Machine-learning models trained directly on electronic spectra can predict spin--orbit-driven properties, yet models with comparable predictive accuracy may learn different electronic dependencies. 
We examine this problem using magnetic anisotropy in atomic-scale magnets and a validation strategy designed to assess the physical relevance of learned spectral relationships. 
We compare a bidirectional gated recurrent unit and a one-dimensional convolutional neural network, both trained to predict magnetic anisotropy energy (MAE) from spin- and orbital-resolved scalar-relativistic densities of states (SR-DOS). 
Despite comparable in-domain (ID) accuracy, the architectures learn only partly overlapping spectral dependencies. 
Shapley additive explanations identify spectral features shared between the two architectures that can be associated with plausible spin--orbit-coupling pathways consistent with second-order perturbation theory (PT2). 
Adding PT2-derived MAE contributions to the model inputs has little effect on ID performance but can improve transfer beyond the training domain, with gains differing between architectures. 
Controlled spectral perturbations further demonstrate that similar attribution patterns do not imply the same functional dependence of the predicted MAE on spectral weight. 
The contrasting responses provide a functional context for architecture-dependent out-of-domain transfer. 
End-to-end learning can identify candidate electronic signatures of magnetic anisotropy. Their credible microscopic interpretation, however, rests on convergent evidence from predictive performance, model comparison, physical theory, controlled spectral interventions, and evaluation under distribution shift.\\
\end{abstract}




\section{Introduction}

Machine-learning (ML) models increasingly use energy-resolved electronic-structure representations to predict materials properties~\cite{Suzuki2021,Fung2021,Mok2023,Takano2026}. 
In particular, neural networks can learn predictive spectral features directly from densities of states (DOS), reducing reliance on handcrafted descriptors~\cite{Fung2021,Mok2023}. 
While the end-to-end approach increases model flexibility, it also makes the physical basis of the predictions more difficult to establish.

Feature-importance methods may identify the energy regions contributing most strongly to a prediction, but the resulting attributions remain properties of the fitted model and cannot be assumed to represent model-independent physical observables~\cite{Holzinger2022,Graham2023,Hwang2025}. This issue is closely related to the Rashomon effect, whereby similarly accurate models may rely on distinct functional dependencies and yield different explanations~\cite{Koutra2023,Laberge2021,Seipel2023,Leonardis2025}. Although multi-model and Rashomon-set analyses have been applied to descriptor-based materials tasks~\cite{Li2023,Li2023a,Li2025}, 
high-dimensional electronic spectra pose an additional challenge. 
Spectral attribution must be evaluated against both the functional response of the model and the underlying physics.


Magnetic anisotropy energy ($\rm MAE$) is central to the stability and control of atomic-scale magnets (ASMs), enabling single-atom information storage, spin manipulation, and quantum functionality~\cite{Gambardella2003,Natterer2017,Yang2019b,Wang2023}. 
Its sensitivity to subtle changes in orbital occupation, hybridization, and the energetic alignment of occupied and unoccupied electronic states makes $\rm MAE$ a demanding test case for assessing the credibility of spectral relationships learned by ML models. 

Recent computational studies have established transition-metal (TM) dimers anchored at nitrogen-decorated single vacancies (NSVs) in graphene, including MgO-supported configurations, as 
well-defined computational platform for studying atomic-scale magnetic anisotropy~\cite{Navratil2025, Navratil2026}.
Extending such investigations to broader chemical, structural, and substrate environments requires costly geometry optimization and noncollinear density functional theory (DFT) calculations including spin--orbit coupling (NCL--SOC). 
Beyond the computational cost, 
the resulting high-dimensional electronic-structure data complicate the identification of the microscopic origins of $\rm MAE$. 

For ASMs, second-order perturbation theory (PT2) provides a physical rationale for using scalar-relativistic densities of states (SR-DOS) as inputs to ML models. 
Within PT2, $\rm MAE$ is related to spin--orbit-mediated couplings between occupied and unoccupied states, subject to symmetry-dependent selection rules~\cite{Wang1993,Hu2014}. Consequently, SR-DOS contains information relevant to $\rm MAE$ prediction without requiring explicit NCL--SOC calculations at inference time.
The perturbative description becomes less reliable when near-degenerate transitions involve energy separations comparable to the SOC strength~\cite{Cinal2026}.
Handcrafted descriptors based on selected peaks, integrals, or fitted components~\cite{Navratil2025, Ishikawa2024} may also omit relevant fine structure. 
Learning directly from SR-DOS preserves the energy-resolved input without imposing a predefined descriptor form.

Here, we introduce a strategy for assessing the physical relevance of learned spectral relationships. 
To distinguish shared dependencies from those specific to model architecture, we compare a bidirectional gated recurrent unit (BiGRU) with a one-dimensional convolutional neural network (1D-CNN). 
Both architectures are trained on spin- and orbital-resolved SR-DOS of TM dimers anchored at NSVs in graphene. 
Single- and multi-task settings are considered, with $\rm MAE$ as the primary target and selected SOC-related quantities as auxiliary targets.
SR-DOS and Shapley additive explanations (SHAP)~\cite{Shapley1953,Lundberg2017} are used to relate spectral features shared between the architectures to SOC pathways expected from PT2. 
Controlled spectral perturbations then test whether the attributed regions are functionally relevant to the predicted $\rm MAE$. 
The architecture-dependent responses are considered alongside differences in out-of-domain (OOD) transfer, evaluated on explicit-substrate systems excluded from training. 
We also examine whether supplying PT2-derived MAE contributions as additional model inputs improves in-domain (ID) prediction or OOD transfer.

Model comparison tests whether identified dependencies are shared between architectures, PT2 provides a physical reference for their interpretation, spectral interventions test their functional role, and OOD evaluation probes their robustness beyond the training distribution.
Together, these analyses provide convergent evidence for assessing whether the learned spectral relationships support a physically grounded microscopic interpretation.


\section{Methods}


\subsection{Machine learning}

A bidirectional gated recurrent unit (BiGRU) and a one-dimensional convolutional neural network (1D-CNN) are used to predict magnetic anisotropy energy ($\rm MAE$), atom-resolved spin--orbit coupling energies ($E_{\rm SOC}$), and orbital-resolved SOC matrix elements ($H_{\rm SOC}^{d}$) from spin- and orbital-resolved scalar-relativistic density of states (SR-DOS). Calculating the SR-DOS is substantially less expensive than the noncollinear density functional theory (NCL--DFT) calculations including SOC used to obtain the target quantities.

The SR-DOS is represented on a uniform grid of $200$ energy points spanning $[-3,3]$~eV relative to the Fermi level ($E_{\rm F}$) using either five $d$-orbital channels or all nine $s$-, $p$-, and $d$-orbital channels for each spin projection and transition-metal (TM) atom.
Before being passed to the models, the SR-DOS is reorganized using a Fermi-level split. The occupied and unoccupied halves are ordered by increasing distance from $E_{\rm F}$ and paired at corresponding sequence positions, with their DOS values retained as separate channels. This representation yields a sequence of length $T=100$, in which each position contains spectral information from both sides of $E_{\rm F}$. 
With the distance from $E_{\rm F}$ included as an additional channel, the input tensor has dimensions $T\times C$, where $C=41$ for the $d$-only representation and $C=73$ for the $s\,p\,d$ representation.
This transformation changes only the structural layout of the SR-DOS grid, facilitating the learning of spectral relationships around $E_{\rm F}$, while preserving the raw computational spectrum. 


\subsubsection*{BiGRU architecture}

The recurrent architecture combines two self-attention modules with a BiGRU encoder. 
The input first passes through a channel self-attention module that captures interactions among the atom-, spin-, and orbital-resolved DOS channels before sequential encoding. Channel embeddings are averaged over the sequence and projected to queries, keys, and values with a projection dimension of $32$. Scaled dot-product attention produces a $C\times C$ channel-mixing matrix that is applied uniformly at every sequence position. The resulting representation is combined with the original input through a residual connection followed by layer normalization.

The attention-weighted sequence is then encoded by a three-layer BiGRU with $100$ hidden units per direction and dropout $p=0.3$ between recurrent layers. The forward and backward hidden states are concatenated, yielding a $200$-dimensional representation at each sequence position. Temporal self-attention applies full query, key, and value projections of dimension $200$ to the BiGRU states and assigns learned weights to positions along the reordered energy sequence. 
The BiGRU output is normalized with layer normalization before being passed to the temporal attention module. Mean pooling over the attention-weighted sequence produces a  single context vector.
Final predictions are generated by two fully connected layers with dimensions $200\rightarrow50\rightarrow n_{\rm out}$ and dropout $p=0.3$, where $n_{\rm out}$ depends on the selected single- or multi-task target configuration. 
The effects of Fermi-level reordering, recurrent directionality, and attention design are evaluated through an ablation study (Table~S1). 
The tested variants include unidirectional and bidirectional GRUs, representations with and without Fermi-level reordering, and channel, temporal, combined channel--temporal, or no explicit attention. 
Fermi-level reordering consistently improves $\rm MAE$ prediction in all three target configurations compared with the corresponding bidirectional models without reordering. Bidirectional encoding generally outperforms the unidirectional alternative, although the magnitude of the improvement depends on the predicted quantity. 
The optimal attention design also depends on the target configuration but combined channel and temporal attention yields the highest $\rm MAE$ $R^2$ in each evaluated setting. 
This configuration is adopted for all BiGRU analyses.


\subsubsection*{1D-CNN architecture}

The convolutional architecture treats the Fermi-reordered SR-DOS as a multichannel signal defined along the distance-from-$E_{\rm F}$ axis. It consists of four successive convolutional blocks, each comprising a one-dimensional (1D) convolution with kernel size $5$, padding $2$, and $100$ output channels, followed by batch normalization, ReLU activation, and dropout $p=0.3$.
The channel width remains fixed throughout the network, and no pooling or strided convolution is applied within the convolutional stack. The sequence length thus remains $T=100$, preserving the resolution of the reordered input representation.

The fixed-width convolutional kernels define local receptive fields along the reordered energy axis, enabling the extraction of local spectral motifs from the SR-DOS. 
Global average pooling over the sequence produces a $100$-dimensional representation. Final predictions are generated by two fully connected layers with dimensions $100\rightarrow100\rightarrow n_{\rm out}$ and dropout $p=0.3$, where $n_{\rm out}$ depends on the selected single- or multi-task target configuration. 

The final convolutional architecture is selected by comparing plain 1D-CNNs of different depths with 1D variants of VGG, ResNet, and DenseNet (Table~S2). 
The four-block 1D-CNN achieves the highest $\rm MAE$ $R^2$ in the ${\rm MAE}+E_{\rm SOC}$ setting and remains close to the best-performing architectures in the other tasks, prompting its use in all 1D-CNN analyses.


\subsubsection*{Training procedure}
Both architectures are trained using the Adam optimiser~\cite{kingma2014adam} with an initial learning rate of $10^{-3}$, reduced by a factor of $0.9$ every $100$ epochs. 
The models minimize the mean squared error (MSE) between predicted and target values. 
Before computing the MSE loss, each target is standardized to zero mean and unit variance ($z$-score normalization) using the mean and standard deviation of the corresponding training partition.

In-domain performance is evaluated using eight-fold cross-validation, with 
the implicit-substrate systems randomly assigned to eight mutually exclusive subsets. In each fold, six subsets are used for training, one for validation, and one as an independent test set.
Early stopping is applied by monitoring the validation loss after each epoch.
If the loss does not improve for $500$ consecutive epochs, training is terminated and the checkpoint with the lowest validation loss is retained. 
The reported $R^2$ scores and mean absolute errors (MAErr) are evaluated on the held-out test subset in each fold and averaged over the eight folds. 
Fold-level variability and confidence intervals for the performance metrics are reported in Section~S2 of the Supplementary Information.
The corresponding training and validation loss curves for all folds and target configurations are provided in Section~S3 of the Supplementary Information.

For the SHAP analysis and out-of-domain evaluation, the models are retrained on the complete ID dataset without a held-out subset. Because early stopping cannot be applied without a validation subset, the number of training epochs is fixed to the mean early-stopping epoch obtained from the eight cross-validation folds for the corresponding  architecture and target configuration. 
The resulting training duration ranges from approximately $750$ to $1900$ epochs.
The explicit-substrate dataset is used exclusively for OOD evaluation and does not contribute to model training, hyperparameter selection, or determination of the training duration.


\subsection{Electronic-structure calculations}

Spin-polarized density functional theory (DFT) calculations are performed using the Vienna \textit{ab-initio} simulation package (VASP)~\cite{Kresse1993, Kresse1994, Kresse1996a, Kresse1996}. 
Exchange and correlation are described within the generalized gradient approximation (GGA)~\cite{Perdew1992, Perdew1992erratum} using the Perdew-Burke-Ernzerhof (PBE) functional~\cite{Perdew1996, Perdew1996aerratum}, and electron-ion interactions are treated using the projector augmented-wave (PAW) method~\cite{Kresse1999, Blochl1994a}.
A plane-wave cutoff energy of $400$~eV is used for both SR and NCL--SOC calculations, following previously established convergence criteria for TM dimers anchored at vacancy defects in graphene~\cite{Navratil2021,Navratil2022}.
Partial occupancies are treated using Gaussian smearing with a width of $0.02$~eV.
Electronic minimization is considered converged when the change in total energy falls below 10$^{-6}$~eV. Symmetry constraints are disabled in all calculations.

Scalar-relativistic structural relaxations are performed using a $2\times2\times1$ $k$-point mesh until the residual forces fall below $10^{-2}$~eV\,\AA$^{-1}$, followed by static SR calculations using $6\times6\times1$ sampling. 
The more demanding NCL--SOC calculations are performed using a $3\times3\times1$ mesh. Selected high-MAE systems are recalculated at $6\times6\times1$, resulting in only minor changes.

The SR-DOS serves as the model input, whereas NCL-DFT calculations including SOC provide the reference targets. 
The ID training dataset comprises 806 OsPd, OsPt, IrPd, and IrPt dimer configurations on freestanding NSV-graphene and on NSV-graphene retaining the geometry induced by Ir(111), as shown in Figure~S2, without explicit substrate atoms~\cite{Navratil2025}. 
OOD transfer is evaluated on an explicit-substrate dataset comprising ground-state configurations on Cu(111), Ni(111), Ir(111), and MgO(001).

The self-consistent 
${\rm MAE}$ is obtained from the NCL--SOC total energies corresponding to different magnetization directions 
and is defined as 
\begin{equation} 
{\rm MAE} = \min\left(E_x,E_{xy},E_y\right)-E_z, 
\end{equation} 
where $E_\alpha$ denotes the total energy for magnetization oriented along direction $\alpha$~\cite{Kleinman1980,MacDonald1980,Hobbs2000,Marsman2002}. 
The $z$ axis is perpendicular to the graphene layer and parallel to the TM--TM bond, 
while $x$, $y$, and $xy$ denote the considered in-plane orientations. 
Under this convention, positive ${\rm MAE}$ corresponds to an out-of-plane easy axis, whereas negative values indicate an in-plane easy axis.

PT2-derived contributions to the MAE are calculated from the energy separations of occupied--unoccupied $d$-orbital pairs 
using the orbital selection rules and the procedure described in Ref.~\cite{Navratil2025}. 
The resulting values are included as an additional input channel in the corresponding physics-informed model variants.


\section{Results}


\subsection{Spectral learning of interdependent spin--orbit-driven properties}

Spin- and orbital-resolved SR-DOS for TM dimers anchored at NSVs in graphene serve as model inputs, and the target quantities are obtained from NCL-DFT calculations including SOC. 
Single- and multi-task models are trained to predict $\rm MAE$, atom-resolved 
$E_{\rm SOC}$, and 16 $d$-orbital SOC matrix elements ($H_{\rm SOC}^{d}$), with all outputs in a given multi-task configuration predicted from a shared spectral representation.
The BiGRU and 1D-CNN models are evaluated using different learning settings and input orbital channels in ID and OOD regimes (cf. Tables~\ref{tab:performance_rnn} and \ref{tab:performance_cnn}).


\begin{table}[htbp]
\centering
\caption{In-domain and out-of-domain performance of the BiGRU models.
The ID data include freestanding and implicit-substrate systems, while the OOD set comprise explicit-substrate systems excluded from training.
The coefficient of determination ($R^2$) and mean absolute error (MAErr, in meV) are provided for single- and multi-task models using either $d$ or $s\,p\,d$ orbital representations. Metrics for $E_{\rm SOC}$ are averaged over the two TM atoms in each dimer, whereas those for $H_{\rm SOC}^{d}$ are additionally averaged over the 16 orbital-pair elements. Values in parentheses indicate the $R^2$ scores obtained without Fermi-level reordering of the BiGRU input sequence.}
\label{tab:performance_rnn}
\begin{tabular}{llcccccc}
\toprule
& & \multicolumn{2}{c}{\textbf{MAE}} & \multicolumn{2}{c}{$\bm{E}_{\mathbf{SOC}}$} & \multicolumn{2}{c}{$\bm{H}_{\mathbf{SOC}}^{\mathbf{d}}$} \\
\cmidrule(lr){3-4} \cmidrule(lr){5-6} \cmidrule(lr){7-8}
\textbf{Model Target} & \textbf{Orbitals} & \textbf{R$^2$} & \textbf{MAErr} & \textbf{R$^2$} & \textbf{MAErr} & \textbf{R$^2$} & \textbf{MAErr} \\
\midrule
\multicolumn{8}{l}{\textbf{ID}} \\
\midrule
$\rm MAE$ & $s\,p\,d$ & 0.804 & 8.84 & -- & -- & -- & -- \\[-0.5ex]
            &         & \multicolumn{1}{c}{\footnotesize (0.753)} & & & & & \\
            & $d$     & 0.786 & 9.26 & -- & -- & -- & -- \\
\addlinespace
${\rm MAE}+E_{\rm SOC}$ & $s\,p\,d$ & 0.821 & 8.26 & 0.867 & 10.7 & -- & -- \\[-0.5ex]
                          &         & \multicolumn{1}{c}{\footnotesize (0.805)} & & & & & \\
                          & $d$     & 0.817 & 8.32 & 0.867 & 11.0 & -- & -- \\
\addlinespace
${\rm MAE}+E_{\rm SOC}$ & $s\,p\,d$ & 0.796 & 9.14 & 0.850 & 12.1 & 0.567 & 4.07 \\[-0.5ex]
$+\, H_{\rm SOC}^{d}$     &         & \multicolumn{1}{c}{\footnotesize (0.726)} & & & & & \\
                          & $d$     & 0.786 & 9.38 & 0.844 & 12.3 & 0.572 & 4.09 \\
\midrule
\multicolumn{8}{l}{\textbf{OOD}} \\
\midrule
$\rm MAE$ & $s\,p\,d$ & 0.591 & 15.9 & -- & -- & -- & -- \\
            & $d$     & 0.450 & 20.7 & -- & -- & -- & -- \\
\addlinespace
${\rm MAE}+E_{\rm SOC}$ & $s\,p\,d$ & 0.771 & 13.7 & 0.608 & 22.6 & -- & -- \\
                          & $d$     & 0.531 & 20.0 & 0.096 & 30.4 & -- & -- \\
\addlinespace
${\rm MAE}+E_{\rm SOC}$ & $s\,p\,d$ & 0.430 & 19.3 & 0.413 & 23.2 & $-0.742$ & 7.68 \\
$+\, H_{\rm SOC}^{d}$     & $d$     & 0.521 & 20.6 & 0.250 & 33.8 & $-0.999$ & 7.78 \\
\bottomrule
\end{tabular}
\end{table}


\begin{table}[h!]
\centering
\caption{In-domain and out-of-domain performance of the 1D-CNN models. 
The coefficient of determination ($R^2$) and mean absolute error (MAErr, in meV) are provided for single- and multi-task models using either $d$ or $s\,p\,d$ orbital representations. Metrics for $E_{\rm SOC}$ are averaged over the two TM atoms in each dimer, whereas those for $H_{\rm SOC}^{d}$ are additionally averaged over the 16 orbital-pair elements.}
\label{tab:performance_cnn}
\begin{tabular}{llcccccc}
\toprule
& & \multicolumn{2}{c}{\textbf{MAE}} & \multicolumn{2}{c}{$\bm{E}_{\mathbf{SOC}}$} & \multicolumn{2}{c}{$\bm{H}_{\mathbf{SOC}}^{\mathbf{d}}$} \\
\cmidrule(lr){3-4} \cmidrule(lr){5-6} \cmidrule(lr){7-8}
\textbf{Model Target} & \textbf{Orbitals} & \textbf{R$^2$} & \textbf{MAErr} & \textbf{R$^2$} & \textbf{MAErr} & \textbf{R$^2$} & \textbf{MAErr} \\
\midrule
\multicolumn{8}{l}{\textbf{ID}} \\
\midrule
$\rm MAE$ & $s\,p\,d$ & 0.808 & 8.38 & -- & -- & -- & -- \\
    & $d$             & 0.801 & 8.69 & -- & -- & -- & -- \\
\addlinespace
${\rm MAE}+E_{\rm SOC}$ & $s\,p\,d$ & 0.842 & 7.71 & 0.876 & 10.4 & -- & -- \\
                          & $d$     & 0.821 & 8.35 & 0.875 & 10.8 & -- & -- \\
\addlinespace
${\rm MAE}+E_{\rm SOC}$ & $s\,p\,d$ & 0.804 & 8.57 & 0.862 & 11.3 & 0.601 & 3.85 \\
$+\, H_{\rm SOC}^{d}$ & $d$         & 0.790 & 9.09 & 0.853 & 11.8 & 0.596 & 3.91 \\
\midrule
\multicolumn{8}{l}{\textbf{OOD}} \\
\midrule
$\rm MAE$ & $s\,p\,d$ & 0.408 & 23.7 & -- & -- & -- & -- \\
            & $d$     & 0.475 & 21.3 & -- & -- & -- & -- \\
\addlinespace
${\rm MAE}+E_{\rm SOC}$ & $s\,p\,d$ & 0.510 & 19.5 & 0.557 & 25.6 & -- & -- \\
                          & $d$     & 0.461 & 20.5 & 0.622 & 25.2 & -- & -- \\
\addlinespace
${\rm MAE}+E_{\rm SOC}$ & $s\,p\,d$ & 0.578 & 20.2 & 0.677 & 23.6 & \phantom{$-$}0.015 & 6.11 \\
$+\, H_{\rm SOC}^{d}$     & $d$     & 0.470 & 20.1 & 0.680 & 24.2 & $-0.093$ & 5.96 \\
\bottomrule
\end{tabular}
\end{table}

\begin{figure}[h!]
\centering
\includegraphics[width=0.49\textwidth]{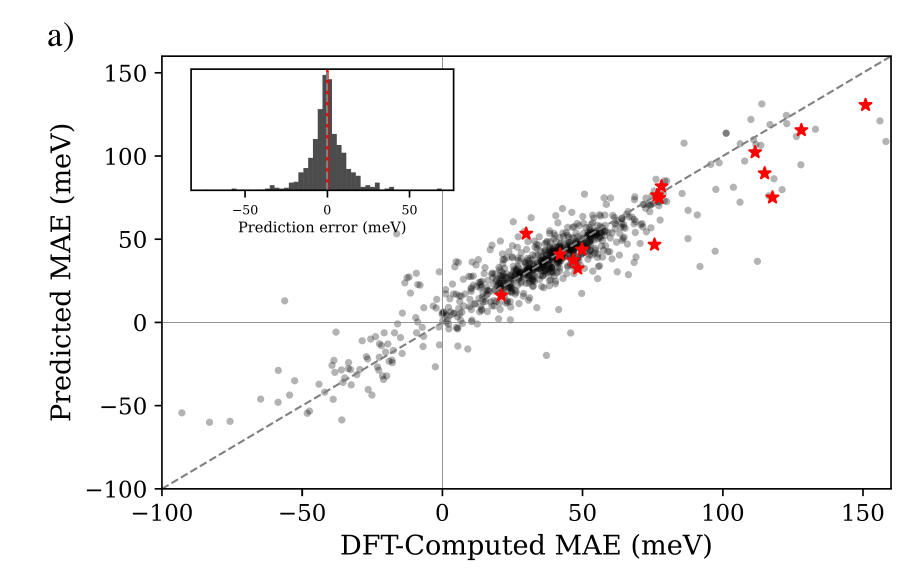}
\includegraphics[width=0.49\textwidth]{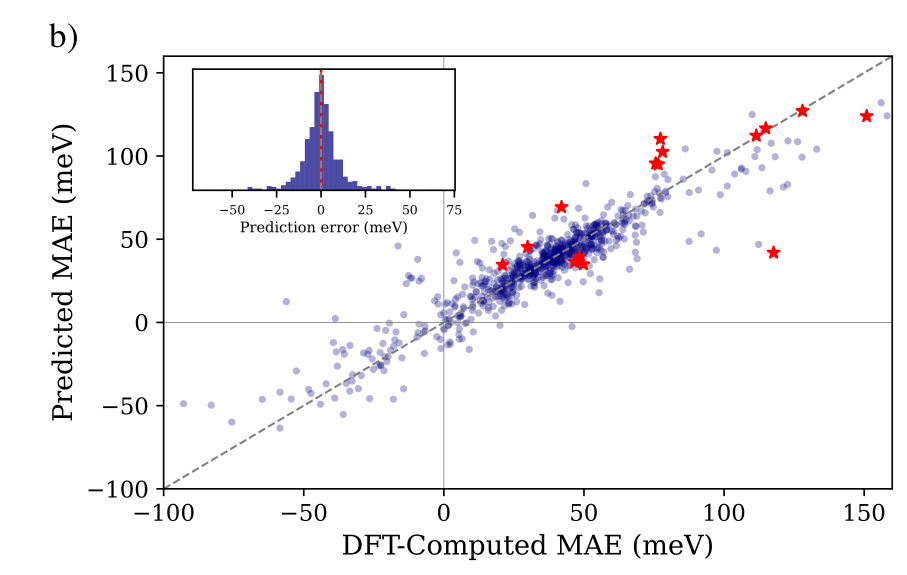}
\caption{Comparison of ML-predicted and DFT-computed magnetic anisotropy energy (MAE).
(a) Predictions from the BiGRU, 
(b) predictions from the 1D-CNN. Both models were trained in the ${\rm MAE}+E_{\rm SOC}$ multi-task setting using the full $s\,p\,d$ SR-DOS representation. 
The gray dashed diagonal indicates perfect agreement with the DFT values. Red stars denote the explicit-substrate systems.}
\label{fig:mae_model}
\end{figure}
\begin{figure}[h]
\centering
\includegraphics[width=\textwidth]{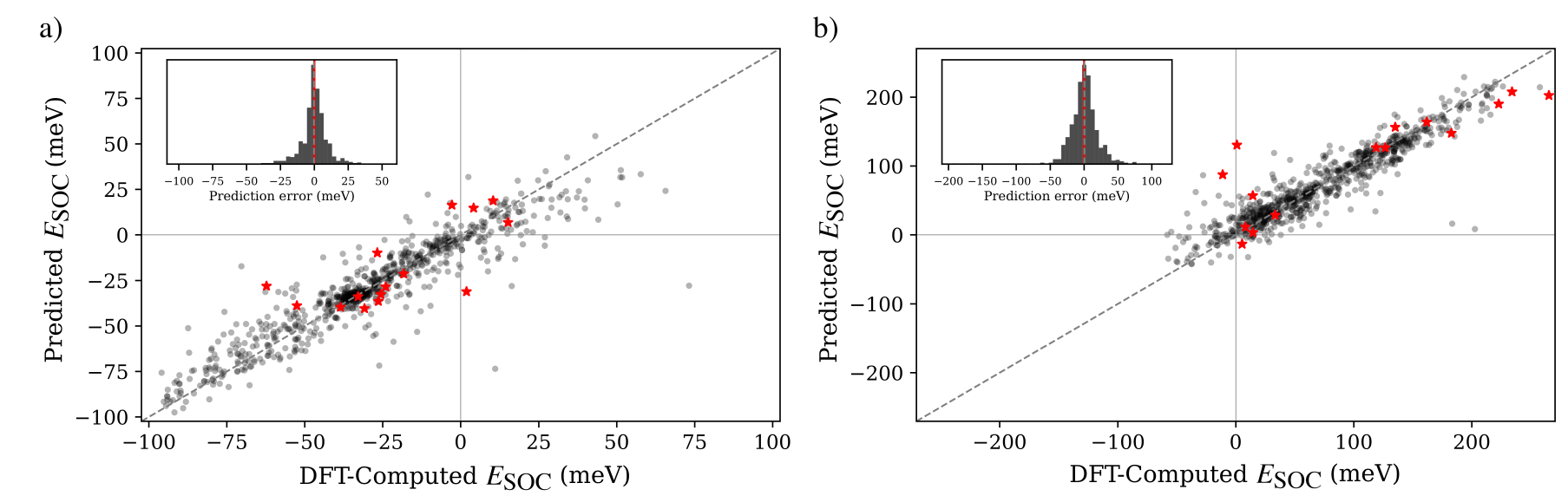}
\includegraphics[width=\textwidth]{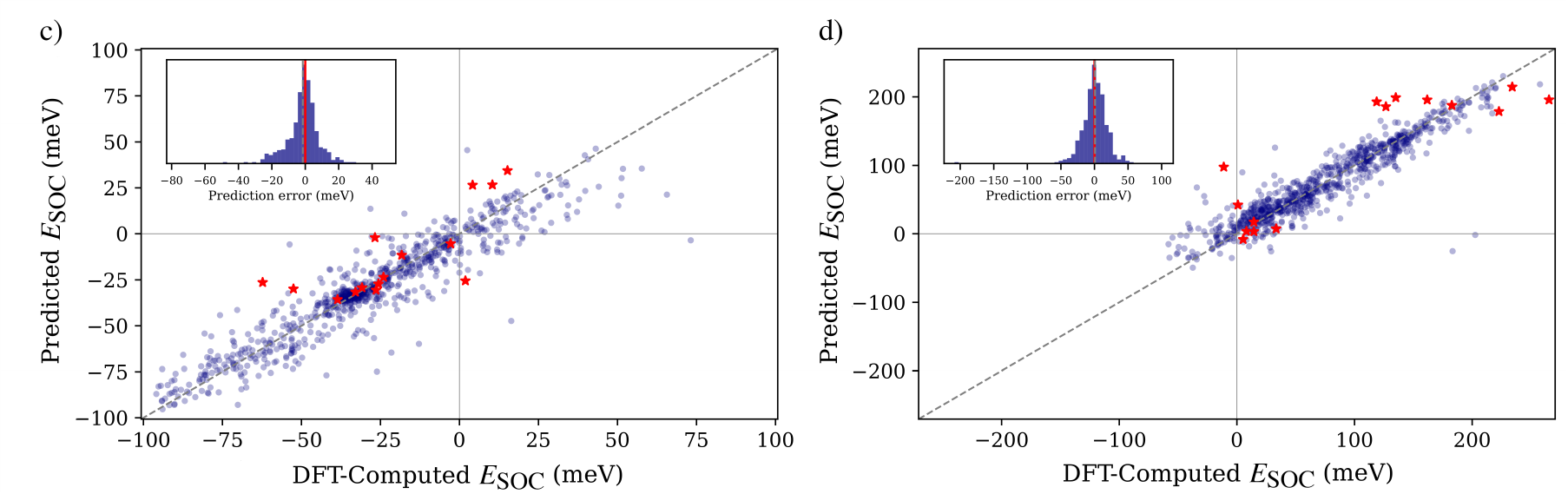}
\caption{Comparison of ML-predicted and DFT-computed atom-resolved SOC energies (${E}_{\rm{SOC}}$). 
BiGRU predictions for (a) the bottom and (b) top TM atoms, 
(c) and (d) corresponding 1D-CNN predictions. Both models were trained in the ${\rm MAE}+E_{\rm SOC}$ multi-task setting using the full $s\,p\,d$ SR-DOS representation. 
The gray dashed diagonal indicates perfect agreement with the DFT values. Red stars denote the explicit-substrate systems.}\label{fig:esoc_model}
\end{figure}

In the ID setting, both architectures achieve comparable predictive performance with a slight overall edge for the 1D-CNN.
Predicting $\rm MAE$ jointly with $E_{\rm SOC}$ consistently improves $\rm MAE$ accuracy relative to the single-target baseline for both architectures and input representations, 
consistent with atom-resolved SOC energies providing useful, physically related auxiliary supervision.
Moreover, $E_{\rm SOC}$ consistently reaches higher $R^2$ values than $\rm MAE$, indicating that its atom-resolved variation is more readily inferred from SR-DOS than the subtle balance of competing SOC contributions determining $\rm MAE$.
The corresponding parity plots for $\rm MAE$ (Figure~\ref{fig:mae_model}) and $E_{\rm SOC}$ (Figure~\ref{fig:esoc_model}) show predictions clustered around the diagonal, with only a small number of more pronounced outliers.


Including the orbital-resolved SOC matrix elements  as additional targets only modestly 
reduces $\rm MAE$ and $E_{\rm SOC}$ accuracy, yet the individual matrix elements remain considerably harder to infer from SR-DOS.
Their $R^2$ values span a broad range, from $0.382$ to $0.844$ for the BiGRU and from $0.375$ to $0.885$ for the 1D-CNN, suggesting that the strength of the mapping between the spectral input and specific orbital couplings varies considerably.

Restricting the spectral representation to the $d$-orbital DOS leads to only small changes in ID performance, consistent with the dominant role of $d$ states in the magnetic anisotropy of ASMs and indicating that the $d$-orbital DOS provides most of the information used for ID prediction. 

The transition to explicit substrates changes the spectral environment presented to the models by introducing substrate-derived states and modifying the hybridization and energetic placement of dimer-related features. OOD transfer consequently tests whether the learned spectral dependencies remain useful under this chemically and electronically distinct environment.

The OOD performance depends more strongly on architecture, orbital representation, and target setting. 
For $\rm MAE$, the highest transfer accuracy is achieved by the $s\,p\,d$ BiGRU trained jointly on $\rm MAE$ and $E_{\rm SOC}$. This model also outperforms the feature-engineered gradient-boosting approach evaluated on the same explicit-substrate benchmark~\cite{Navratil2025}. 
Extending the target set to include $H_{\rm SOC}^{d}$ substantially degrades $\rm MAE$ transfer for the $s\,p\,d$ BiGRU, while producing little change for its $d$-only variant. 
By comparison, the OOD $\rm MAE$ $R^2$ of the 1D-CNN increases for both orbital representations, with the larger improvement obtained for the full $s\,p\,d$ input.

For $E_{\rm SOC}$ in the two-target setting, the BiGRU benefits markedly from the full $s\,p\,d$ representation, whereas the $d$-only representation yields better transfer for the 1D-CNN. 
In the full multi-task setting, the 1D-CNN retains substantially higher OOD $E_{\rm SOC}$ accuracy than the BiGRU. 
The orbital-resolved matrix elements themselves do not transfer reliably in either architecture. Their OOD $R^2$ values are strongly negative for the BiGRU and remain close to zero for the 1D-CNN.

Comparable ID accuracy alongside architecture- and target-dependent OOD MAE transfer motivates a comparison of how the two models use the same $d$-orbital spectral representation.


\subsection{Energy-resolved SHAP analysis}

To identify the SR-DOS features associated with the learned $\rm MAE$ predictions, we analyze energy-resolved SHAP attributions for the BiGRU and 1D-CNN, as detailed in Section~S5 of the Supplementary Information.
Spectral regions highlighted by both architectures are treated as candidate $\rm MAE$-relevant signatures that are less readily attributed to a single architecture-specific strategy.
SHAP values are computed relative to a background distribution $\mathcal{B}$ defined by low-anisotropy systems  with $\rm MAE \in [-25,\,25]~{\rm meV}$. 
Positive and negative SHAP values indicate spectral contributions that increase or decrease the predicted $\rm MAE$ relative to $\mathcal{B}$, whereas the absolute SHAP magnitude quantifies the attributed importance of a given energy region independently of sign.

\begin{figure}[h]
\centering
\includegraphics[width=0.49\textwidth]{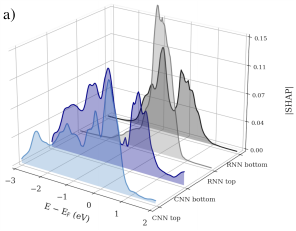}
\includegraphics[width=0.49\textwidth]{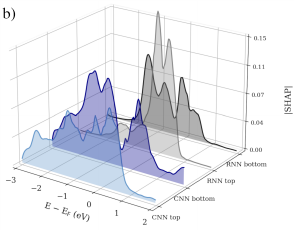} \\
\includegraphics[width=0.49\textwidth]{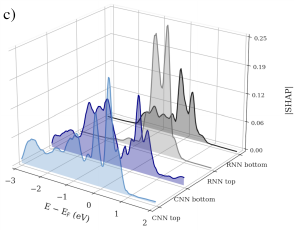}
\includegraphics[width=0.49\textwidth]{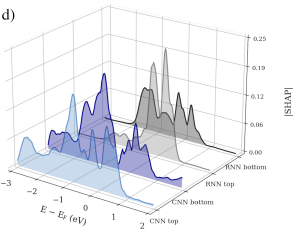} 
\caption{Energy-resolved average absolute SHAP profiles for the top and bottom TM atoms. 
(a, b) High-positive-MAE systems (${\rm MAE}>25$~meV), 
(c, d) strongly negative-MAE systems (${\rm MAE}< -25$~meV). 
Results were obtained from models trained on the $d$-orbital SR-DOS representation using the $\rm MAE$-only target in (a) and (c), and the ${\rm MAE}+E_{\rm SOC}$ target setting in (b) and (d). 
The SHAP background distribution is defined by weak-anisotropy systems with $-25~{\rm meV}\leq{\rm MAE}\leq25~{\rm meV}$.
} 
\label{fig:esoc_avg_shap}
\end{figure}

For both high-positive- and strongly negative-$\rm MAE$ systems, the two architectures consistently assign attribution to Fermi-adjacent $d$-state regions, making them the most reproducible spectral signature shared by the models (Figure~\ref{fig:esoc_avg_shap}). 
The detailed attribution pattern, however, depends on both the network architecture and the prediction target.
For the top TM atom, the BiGRU gives a stronger response immediately adjacent to $E_{\rm F}$. Below these features, the 1D-CNN response exceeds that of the BiGRU and extends farther into the
occupied region. 
For the bottom TM atom, Fermi-adjacent attribution also recurs, but the BiGRU tends to favor occupied frontier features, whereas the 1D-CNN assigns relatively greater weight above $E_{\rm F}$ and at deeper occupied energies.

Adding $E_{\rm SOC}$ to the target set largely preserves the SHAP-active energy windows while redistributing their relative weight (Figure~\ref{fig:esoc_avg_shap}b,d). 
For the 1D-CNN, a deeper occupied feature on the top TM atom gains importance in both $\rm MAE$ regimes, reaching a magnitude comparable to the Fermi-adjacent peaks for high-positive $\rm MAE$ and becoming the dominant contribution for strongly negative $\rm MAE$. 
For the latter case, a bottom-atom peak emerges at a similar energy. 
For the BiGRU, the main redistribution occurs on the bottom TM atom, where a strong maximum develops deeper in the occupied region. 


\begin{figure}[h!]
\centering
\includegraphics[width=0.95\textwidth]{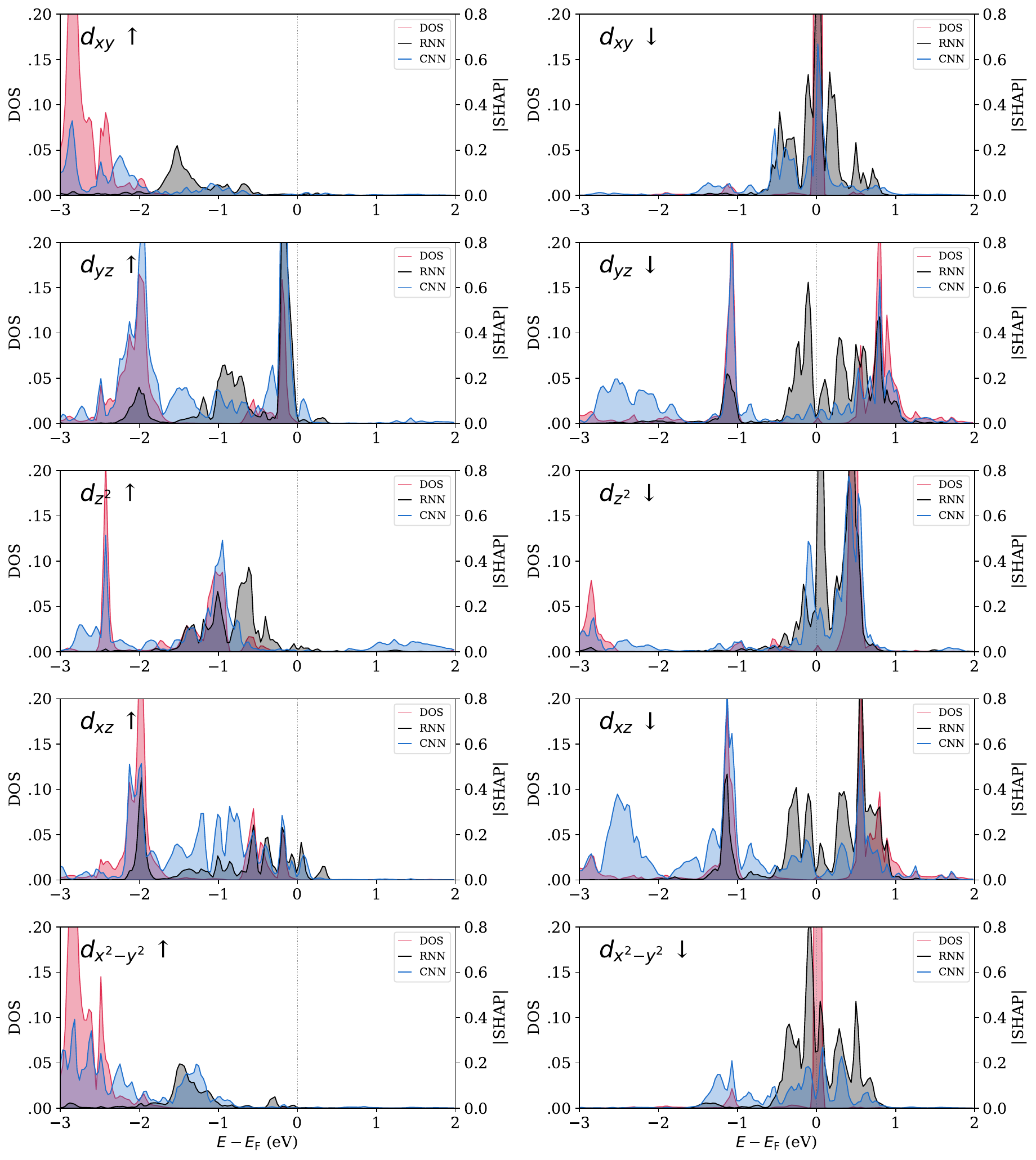}
\caption{Energy-resolved DOS--SHAP overlays. 
The profiles correspond to the top TM atom of the high-MAE OsPd@NSV dimer with a DFT-computed $\rm MAE$ of $133.0$~meV.
Columns and rows distinguish the spin channels and individual $d$ orbitals, respectively. 
}
\label{fig:shap_orbital_highmae}
\end{figure}

Figure~\ref{fig:shap_orbital_highmae} together with Figures~S3--S5 provide atom-, spin-, and orbital-resolved DOS--SHAP views of representative high-positive- and strongly negative-$\rm MAE$ systems. 
The attributions extend beyond the dominant DOS maxima to peak edges, depleted frontier regions, and low-DOS intervals separating occupied and unoccupied spectral features, showing that the resulting patterns are not determined solely by DOS intensity (see also Figures~S6--S11).

For the top TM atom of the highest-$\rm MAE$ dimer (Figures~\ref{fig:shap_orbital_highmae} and~S12), the two architectures highlight overlapping Fermi-adjacent orbital features but weight them differently, with the 1D-CNN retaining attribution farther into the occupied region. 
In the spin-down channel, the $d_{xy}$ and $d_{x^2-y^2}$ SHAP profiles follow prominent Fermi-adjacent DOS features and extend into their depleted surroundings. 
A related peak-edge response appears for $d_{z^2}$, whereas the $d_{xz}$ and $d_{yz}$ channels carry attribution on DOS features flanking $E_{\rm F}$ and within the low-DOS interval separating them. 
By comparison, the spin-up response is predominantly confined to occupied features, albeit with a similar extension beyond the DOS maxima. 

In a PT2 interpretation, the SHAP-active orbital features on the top TM atom are compatible with several symmetry-allowed SOC pathways~\cite{Navratil2025}.
The spin-down $d_{xz}/d_{yz}$ features are consistent with candidate spin-conserving virtual excitations mediated by $\hat{L}_z$.
The $d_{xy}/d_{x^2-y^2}$ pair is likewise $\hat{L}_z$-allowed, although the absence of an occupied counterpart shifts the interpretation of its extended SHAP response from an active virtual-transition channel to the local spectral arrangement of these orbitals.
The $d_{z^2}$-related attribution is consistent with $\hat{L}_x/\hat{L}_y$-mediated coupling to $d_{xz}$/$d_{yz}$, while the $d_{xy}/d_{x^2-y^2}$ response points to additional $\hat{L}_x/\hat{L}_y$-allowed transition channels involving $d_{xz}/d_{yz}$.
Occupied spin-up features indicate spin-flip contributions to the net SOC balance.
For the bottom TM atom, the attribution is distributed over orbital pairs compatible with competing spin-conserving $\hat{L}_z$-allowed and $\hat{L}_x/\hat{L}_y$-mediated channels (Figure~S3), with no single PT2 pathway clearly isolated by the DOS--SHAP pattern.

For the strongly negative-$\rm MAE$ dimer, the dominant top-atom DOS--SHAP features are compatible with same-spin $\hat{L}_x/\hat{L}_y$-mediated coupling between occupied $d_{z^2}$ and $d_{xy}/d_{x^2-y^2}$ states and unoccupied $d_{xz}/d_{yz}$ states (Figures~S4 and~S13) favoring in-plane anisotropy.
For the bottom TM atom (Figure~S5), as in the high-positive-$\rm MAE$ case, the DOS--SHAP pattern reflects competing $\hat{L}_z$- and $\hat{L}_x/\hat{L}_y$-allowed channels.

These orbital assignments provide a PT2-consistent interpretation of the SHAP-active regions. The DOS--SHAP overlays remain model-attribution patterns, however, and do not constitute a direct decomposition of the SOC contribution to MAE.


\section{Discussion}

To test whether SHAP-active spectral regions correspond to functionally important components of the learned mapping, we perform controlled perturbations of the input SR-DOS and recompute the $\rm MAE$ using the trained networks.
The perturbed DOS is constructed as 
\begin{equation} 
DOS_{i,\sigma}^{\rm mod}(E) = \max\left[ DOS_{i,\sigma}(E) + H\exp\left(-\frac{(E-L)^2}{2W^2}\right), 0 \right], 
\end{equation}
where $L$, $H$, and $W$ denote the center, signed amplitude, and width of the perturbation, respectively, with the values of $L$ and $W$ and the maximum value of $H$ for each perturbation summarized in Table~S7. 
Positive values of $H$ add spectral weight, whereas negative values attenuate an existing DOS feature. Existing peaks are selectively reduced, and artificial Gaussian peaks are introduced into depleted energy windows.
The DOS--SHAP overlays (Figure~\ref{fig:shap_orbital_highmae} and Figures~S3--S5) are used to select the spin--orbital channels and energy windows for perturbation, including regions with either strong or negligible SHAP attribution.
The resulting changes in predicted $\rm MAE$ provide an intervention-based test of the learned spectral dependence.

Modifying a single spin- and orbital-resolved DOS channel while keeping all others fixed breaks correlations and sum rules present in self-consistent DFT solutions, potentially affecting the electron count, spin balance, and inter-orbital hybridization. 
Consequently, the modified spectra need not correspond to physically realizable electronic structures and generally fall outside the training distribution. 
The objective is not to generate realistic electronic structures, but to probe the response of the learned mapping to prescribed modifications around representative DOS configurations.

\begin{figure}[h]
\centering
\includegraphics[width=0.95\textwidth]{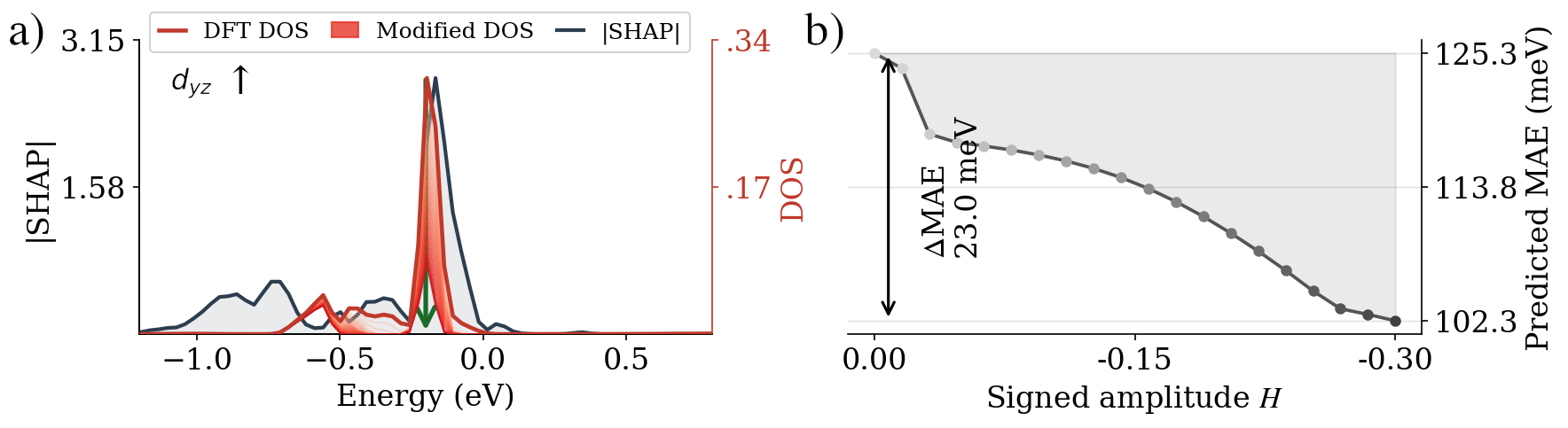} \\
\includegraphics[width=0.95\textwidth]{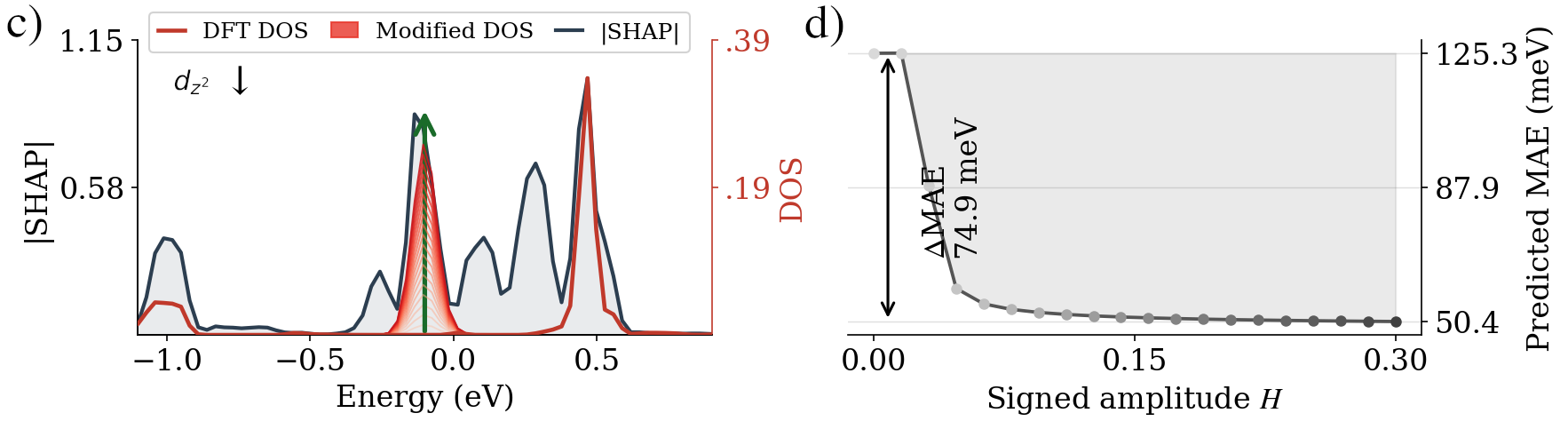} 
\caption{Controlled DOS perturbations evaluated using the BiGRU. 
Results are shown for the bottom TM atom of the high-MAE OsPd@NSV dimer  with a DFT-computed $\rm MAE$ of $133.0$~meV. 
(a) Attenuation of an existing DOS peak in a region with strong SHAP attribution, 
(c) introduction of spectral weight into an initially depleted interval with strong attribution. 
(b, d) Corresponding MAE predictions as functions of the perturbation amplitude $H$.
} \label{fig:shap_dos_mod_highmae}
\end{figure}

For the high-positive-$\rm MAE$ system, attenuating a DOS peak coinciding with a pronounced SHAP maximum substantially lowers the BiGRU-predicted $\rm MAE$ (Figure~\ref{fig:shap_dos_mod_highmae}a,b), indicating a positive contribution of this feature to the prediction. 
Introducing spectral weight into a depleted interval carrying strong SHAP attribution produces an even larger $\rm MAE$ reduction (Figure~\ref{fig:shap_dos_mod_highmae}c,d), revealing that the BiGRU treats the locally depleted DOS as part of the high-$\rm MAE$ spectral signature.  
Perturbations of SHAP-inactive regions have negligible effects (Figure~S14).


The BiGRU response for the strongly negative-$\rm MAE$ system similarly distinguishes SHAP-active from SHAP-inactive regions.
Modifying SHAP-active features shifts the predicted $\rm MAE$ toward less negative values and thus toward reduced in-plane anisotropy (Figure~S15), whereas analogous perturbations in SHAP-inactive intervals leave the prediction nearly unchanged (Figure~S16).
The selected interventions in both MAE regimes thus demonstrate that the BiGRU relies predominantly on spectral features identified by the attribution analysis.

\begin{figure}[h]
\centering
\includegraphics[width=0.95\textwidth]{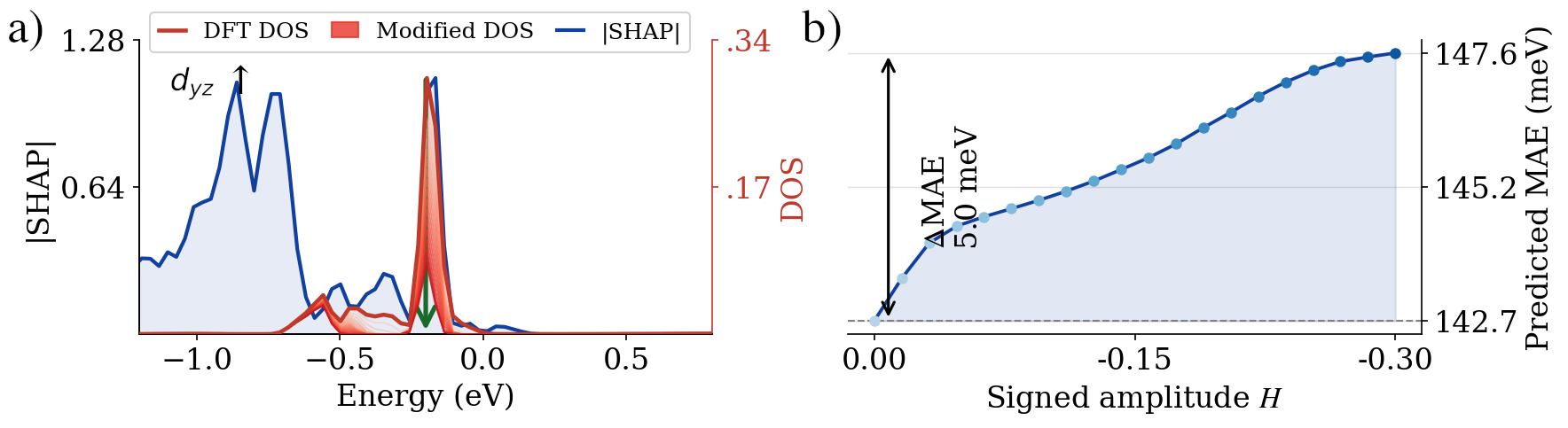} \\
\includegraphics[width=0.95\textwidth]{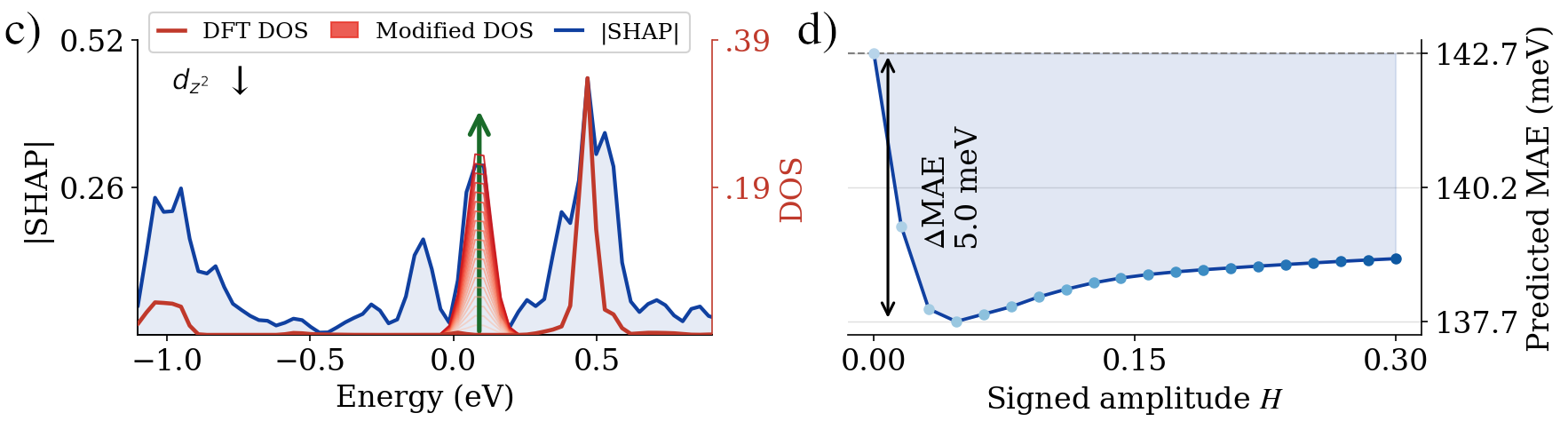} 
\caption{Controlled DOS perturbations evaluated using the 1D-CNN. 
Results are shown for the bottom TM atom of the high-$\rm MAE$ OsPd@NSV dimer with a DFT-computed $\rm MAE$ of $133.0$~meV. 
(a) Attenuation of an existing DOS peak in a region with strong SHAP attribution, 
(c) introduction of spectral weight into an initially depleted interval with strong attribution. 
(b, d) Corresponding MAE predictions as functions of the perturbation amplitude $H$. 
}\label{fig:shap_dos_mod_highmae-CNN}
\end{figure}

The 1D-CNN exhibits a more complex response for the high-positive-$\rm MAE$ system. 
Attenuating the DOS peak that lowers the BiGRU prediction increases the CNN-predicted $\rm MAE$ (Figure~\ref{fig:shap_dos_mod_highmae-CNN}a,b), indicating opposite local responses of the two learned mappings to the same spectral modification.  
Magnetic-force-theorem (MFT)~\cite{Mackintosh1980,Heine1980} analysis performed using POET~\cite{Navratil2026} (Figures~S12 and~S17) provides a plausible physical context for this divergence. The attenuated $d_{yz}$ feature participates in competing contributions to the MAE, including a negative contribution associated with its magnetization-dependent eigenvalue shift and a positive contribution arising from its interplay with the magnetization-dependent occupation of the $d_{xy}/d_{x^2-y^2}$ states. The opposite model responses may reflect different sensitivities to these competing components of the same spectral feature.
Adding spectral weight to the strongly attributed depleted interval lowers the CNN prediction in agreement with the BiGRU response, although the effect is substantially weaker (Figure~\ref{fig:shap_dos_mod_highmae-CNN}c,d). 
Reducing a DOS peak carrying little attribution in the BiGRU but substantial attribution in the 1D-CNN lowers the CNN prediction (Figure~S18a,b), consistent with the CNN SHAP profile.
Unlike the corresponding BiGRU intervention, introducing spectral weight into a window with negligible attribution in the unmodified CNN input produces a pronounced reduction in the CNN-predicted MAE (Figure~S18c,d).
For the strongly negative-$\rm MAE$ system, the 1D-CNN perturbation responses closely agree with those of the BiGRU and preserve the distinction between SHAP-active and SHAP-inactive regions (Figures~S19 and S20).


The close agreement between the two architectures for the strongly negative-$\rm MAE$ system contrasts with their more differentiated behaviors for the high-positive-$\rm MAE$ case, particularly upon introduction of a previously absent spectral motif. 
These system-dependent responses 
provide model-level context for the differences in OOD transfer and motivate examining whether adding PT2-derived MAE contributions to the model input can improve generalization beyond the implicit-substrate domain.

In the ID regime, adding the PT2-derived input channel alters $R^2$ by no more than $0.03$, with the direction of the change depending on both architecture and target setting (Table~\ref{tab:pt2_comparison}). 
The BiGRU improves slightly in the $\rm MAE$-only and ${\rm MAE} + E_{\rm SOC}$ settings but deteriorates when $H_{\rm SOC}^{d}$ is included, whereas the 1D-CNN shows a small benefit only in the full multi-task setting. 
The limited and non-systematic ID effect indicates that the baseline models already capture most of the predictive information provided by the PT2-derived input.


\begin{table}[htbp]
\centering
\caption{Effect of PT2-derived input channel on ID and OOD MAE performance. $R^2$ values are compared for BiGRU and 1D-CNN models with the additional PT2-derived input (w/ PT2) and the corresponding baselines (w/o PT2). Bold values indicate the higher $R^2$ within each comparison pair.}
\label{tab:pt2_comparison}
\begin{tabular}{lcccc}
\toprule
& \multicolumn{2}{c}{\textbf{BiGRU}} & \multicolumn{2}{c}{\textbf{1D-CNN}} \\
\cmidrule(lr){2-3} \cmidrule(lr){4-5}
\textbf{Model Target} & \textbf{w/o PT2} & \textbf{w/ PT2} & \textbf{w/o PT2} & \textbf{w/ PT2} \\
\midrule
\multicolumn{5}{l}{\textbf{ID}} \\
\midrule
$\rm MAE$                               & 0.804 & \textbf{0.812} & \textbf{0.808} & 0.807 \\
${\rm MAE} + E_{\rm SOC}$               & 0.821 & \textbf{0.828} & \textbf{0.842} & 0.831 \\
${\rm MAE} + E_{\rm SOC} + H_{\rm SOC}^d$ & \textbf{0.796} & 0.770 & 0.804 & \textbf{0.812} \\
\midrule
\multicolumn{5}{l}{\textbf{OOD}} \\
\midrule
$\rm MAE$                               & 0.591 & \textbf{0.651} & 0.408 & \textbf{0.523} \\
${\rm MAE} + E_{\rm SOC}$               & \textbf{0.771} & 0.641 & 0.510 & \textbf{0.597} \\
${\rm MAE} + E_{\rm SOC} + H_{\rm SOC}^d$ & 0.430 & \textbf{0.631} & 0.578 & \textbf{0.600} \\
\bottomrule
\end{tabular}
\end{table}


The effect becomes more pronounced under OOD transfer.
The 1D-CNN benefits from the additional PT2-derived input in every target setting, although the magnitude of the improvement varies.
The BiGRU response is less uniform, with PT2 improving the $\rm MAE$-only model and yielding its largest benefit in the full multi-task setting, which additionally includes orbital-resolved SOC targets.
By contrast, the intermediate ${\rm MAE}+E_{\rm SOC}$ model shows a substantial deterioration, further ruling out a universal regularization effect. 
The benefit of the PT2 constraint depends on both architectural bias and the selected combination of prediction targets.


\section{Conclusions}

The combined results place clear limits on what can be inferred from a single post hoc explanation. 
Agreement between the two architectures identifies spectral signatures less readily attributed to a single model bias, whereas the perturbation tests show that overlapping attribution need not imply the same local dependence on spectral weight.   
The PT2 analysis places these signatures within plausible SOC pathways, while 
the distinct OOD profiles show that architecture-dependent behavior extends beyond attribution to predictive transfer.
Finally, including PT2-derived MAE contributions can improve performance in selected OOD settings but does not provide a universally effective route to generalization. 

This work demonstrates that end-to-end learning from raw electronic structure can reveal testable spectral signatures associated with magnetic anisotropy without relying on handcrafted descriptors. 
Assessing their microscopic significance requires convergent evidence from complementary architectures, physical theory, controlled interventions, and OOD evaluation.


\section*{Associated Content}

\section*{Data Availability Statement}

Data supporting the findings of this paper are provided in the
Supporting Information. Both the paper and the Supporting
Information, as well as all data contained therein are freely
available at \url{https://doi.org/10.5281/zenodo.21636936}.


\noindent The source code, trained models, and datasets underlying the reported analyses are available at \url{https://github.com/RafalTopolnicki/ml-dos2mae}.


\section*{Supporting Information}

Architecture and hyperparameter comparison; uncertainty in performance metrics; training curves; structures of the investigated systems; energy-resolved SHAP analyses; DOS--SHAP overlays; comparison of SHAP attributions with magnetic force theorem (MFT) contributions; and additional results from controlled DOS perturbations.


\section*{Author Contributions}

R.T. proposed the research concept, designed, implemented, and evaluated the machine-learning models and the associated attribution and perturbation analyses, and contributed to the interpretation of the model predictions. 
J.N. performed the DFT calculations, prepared the electronic-structure data, and contributed to the analysis and discussion of the results. 
P.B. formulated the research objectives and scope, supervised and coordinated the study, contributed to the analysis and interpretation of the results, and wrote the manuscript.
All authors reviewed and edited the manuscript and approved the final version.   


\section*{Notes}

The authors declare no competing interests.

\section*{Acknowledgments}

This article has been produced with the financial support of the European Union under the REFRESH -- Research Excellence For REgion Sustainability and High-tech Industries project (CZ.10.03.01/00/22\_003/0000048) via the Operational Programme Just Transition and under the TECHSCALE project (CZ.02.01.01/00/22\_008/0004587) via the Operational Programme Johannes Amos Comenius. 

\noindent This work was also supported by the bilateral Czech Science Foundation (GA \v{C}R) / Poland National Science Centre (NCN) project (26-23324K). 

\noindent R.T. acknowledges support from Dioscuri, a programme initiated by the Max Planck Society, jointly managed with the National Science Centre in Poland, and mutually funded by the Polish Ministry of Science and Higher Education and the German Federal Ministry of Research, Technology and Space. 

\noindent J.N. acknowledges support from Palack\'y University Olomouc (project IGA\_PrF\_2026\_002).

\noindent The calculations were performed with the support of the Interdisciplinary Centre for Mathematical and Computational Modeling of the University of Warsaw (ICM UW) under computational grant number g105-2705.


\printbibliography



\section*{Supplementary material}

# Supplementary information:

## Can end-to-end learning from raw electronic structure explain magnetic anisotropy?

Rafał Topolnicki[1,2*], Jan Navrátil[3,4], Piotr Błoński[3,5*]

[1]Institute of Experimental Physics, University of Wrocław, pl. M. Borna 9, 50-204 Wrocław, Poland
[2]Dioscuri Center in Topological Data Analysis, Institute of Mathematics, Polish Academy of Sciences, ul. Śniadeckich 8, 00-656 Warsaw, Poland
[3]Regional Centre of Advanced Technologies and Materials, Czech Advanced Technology and Research Institute (CATRIN), Palacký University Olomouc, Šlechtitelů 27, 779 00 Olomouc, Czech Republic
[4]Department of Physical Chemistry, Faculty of Science, Palacký University Olomouc, 17. listopadu 12, 779 00 Olomouc, Czech Republic
[5]IT4Innovations, VŠB–Technical University of Ostrava, 17. listopadu 2172/15, 708 00 Ostrava-Poruba, Czech Republic

* email: rafal.topolnicki@uwr.edu.pl; piotr.blonski@upol.cz

## Contents

## S1 Architecture and hyperparameter comparison

Tables S1 and S2 compare the in-domain performance of the tested BiGRU and 1D-CNN configurations using eight-fold cross-validation. Each table reports $R^2$ on both the validation set, which is used for early stopping in each fold and architecture selection, and the held-out test set, which is not used in any model selection decision. The close agreement between validation and test $R^2$ across all configurations indicates that the validation-based selection did not lead to overfitting to the validation set. The reported results use the $s\,p\,d$ orbital representation. The selected configurations are also evaluated using the $d$-only representation, for which similar performance trends are observed.

The BiGRU variants differ in Fermi-level reordering, recurrent directionality, and attention design. The evaluated attention mechanisms include channel self-attention, temporal self-attention, their combination, and no explicit attention. The configuration combining Fermi-level reordering, bidirectional encoding, and channel and temporal attention gives the highest MAE $R^2$ in all three target settings, reaching 0.804, 0.821, and 0.796 for the MAE-only, MAE+$E_{\mathrm{SOC}}$, and MAE+$E_{\mathrm{SOC}}$+$H^d_{\mathrm{SOC}}$ settings, respectively. This configuration is used throughout the BiGRU analyses.

The convolutional comparison includes plain 1D-CNN, ResNet, DenseNet, and VGG-style architectures with different numbers of convolutional blocks. Fermi-level reordering is applied to all tested configurations. The four-block plain 1D-CNN gives the highest MAE $R^2$ in the MAE + $E_{\mathrm{SOC}}$ setting, reaching 0.842, and remains within 0.02 of the best-performing architecture in the other target settings. This configuration provides a consistent convolutional backbone across the single- and multi-task comparisons without introducing residual connections or dense interlayer connectivity and is used throughout the 1D-CNN analyses.

**Table S1 | Comparison of BiGRU architecture variants using the *spd* orbital representation.** In-domain $R^2$ values obtained by eightfold cross-validation are reported for different target settings, input arrangements, recurrent directions, and attention mechanisms. *F.-split* indicates whether Fermi-level reordering is applied to the input sequence. *Dir.* denotes a bidirectional (bi) or unidirectional (uni) GRU. *Attention* specifies channel self-attention (ch), temporal self-attention (temp), their combination (ch+temp), or no explicit attention (none). The $E_{\mathrm{SOC}}$ scores are reported separately for the bottom and top TM atoms. Bold entries identify the BiGRU configuration used throughout the analyses, which combines Fermi-level reordering, bidirectional encoding, and channel and temporal attention; this selection was based solely on validation-set performance. The same architecture is used for models trained on the *d*-orbital representation.

| | | | | Validation set | | | Test set | | |
|---|---|---|---|---|---|---|---|---|---|
| **Target** | **F.-split** | **Dir.** | **Attention** | $R^2_{\mathrm{MAE}}$ | $R^2_{E^{\mathrm{bot}}_{\mathrm{SOC}}}$ | $R^2_{E^{\mathrm{top}}_{\mathrm{SOC}}}$ | $R^2_{\mathrm{MAE}}$ | $R^2_{E^{\mathrm{bot}}_{\mathrm{SOC}}}$ | $R^2_{E^{\mathrm{top}}_{\mathrm{SOC}}}$ |
| MAE | no | bi | none | 0.795 | | | 0.753 | | |
| MAE | no | uni | none | 0.783 | | | 0.746 | | |
| MAE | yes | **bi** | **ch+temp** | **0.837** | | | **0.804** | | |
| MAE | yes | bi | channel | 0.818 | | | 0.798 | | |
| MAE | yes | bi | none | 0.815 | | | 0.787 | | |
| MAE | yes | bi | temporal | 0.832 | | | 0.803 | | |
| MAE | yes | uni | none | 0.819 | | | 0.785 | | |
| $\mathrm{MAE} + E_{\mathrm{SOC}}$ | no | bi | none | 0.791 | 0.880 | 0.871 | 0.805 | 0.829 | 0.869 |
| $\mathrm{MAE} + E_{\mathrm{SOC}}$ | no | uni | none | 0.804 | 0.879 | 0.850 | 0.773 | 0.815 | 0.819 |
| $\mathrm{MAE} + E_{\mathrm{SOC}}$ | yes | **bi** | **ch+temp** | **0.812** | **0.891** | **0.891** | **0.821** | **0.857** | **0.877** |
| $\mathrm{MAE} + E_{\mathrm{SOC}}$ | yes | bi | channel | 0.813 | 0.904 | 0.883 | 0.811 | 0.857 | 0.881 |
| $\mathrm{MAE} + E_{\mathrm{SOC}}$ | yes | bi | none | 0.811 | 0.886 | 0.884 | 0.814 | 0.859 | 0.876 |
| $\mathrm{MAE} + E_{\mathrm{SOC}}$ | yes | bi | temporal | 0.817 | 0.889 | 0.896 | 0.804 | 0.865 | 0.898 |
| $\mathrm{MAE} + E_{\mathrm{SOC}}$ | yes | uni | none | 0.806 | 0.886 | 0.880 | 0.806 | 0.848 | 0.882 |
| $\mathrm{MAE} + E_{\mathrm{SOC}} + H^d_{\mathrm{SOC}}$ | no | bi | none | 0.724 | 0.838 | 0.817 | 0.726 | 0.813 | 0.805 |
| $\mathrm{MAE} + E_{\mathrm{SOC}} + H^d_{\mathrm{SOC}}$ | no | uni | none | 0.715 | 0.831 | 0.799 | 0.717 | 0.813 | 0.818 |
| $\mathrm{MAE} + E_{\mathrm{SOC}} + H^d_{\mathrm{SOC}}$ | yes | bi | **ch+temp** | **0.775** | **0.862** | **0.857** | **0.796** | **0.840** | **0.860** |
| $\mathrm{MAE} + E_{\mathrm{SOC}} + H^d_{\mathrm{SOC}}$ | yes | bi | channel | 0.764 | 0.848 | 0.849 | 0.760 | 0.816 | 0.832 |
| $\mathrm{MAE} + E_{\mathrm{SOC}} + H^d_{\mathrm{SOC}}$ | yes | bi | none | 0.768 | 0.860 | 0.843 | 0.767 | 0.838 | 0.840 |
| $\mathrm{MAE} + E_{\mathrm{SOC}} + H^d_{\mathrm{SOC}}$ | yes | bi | temporal | 0.754 | 0.859 | 0.868 | 0.765 | 0.830 | 0.847 |
| $\mathrm{MAE} + E_{\mathrm{SOC}} + H^d_{\mathrm{SOC}}$ | yes | uni | none | 0.773 | 0.850 | 0.838 | 0.771 | 0.830 | 0.848 |

**Table S2 | Comparison of 1D-CNN architecture variants using the $spd$ orbital representation.** In-domain $R^2$ values obtained by eightfold cross-validation are reported for different target settings, convolutional backbones, and network depths. *Arch.* denotes the convolutional backbone, comprising plain CNN, ResNet, DenseNet, and VGG-style architectures. *Depth* specifies the number of convolutional blocks for the plain CNN, ResNet, and VGG-style architectures and the number of dense blocks for DenseNet. The $E_{\mathrm{SOC}}$ scores are reported separately for the bottom and top TM atoms. The bold row identifies the four-block plain 1D-CNN used throughout the analyses; this selection was based solely on validation-set performance. The same architecture is used for models trained on the $d$-only representation. Fermi-level reordering is applied to all configurations.

| | | | Validation set | | | Test set | | |
|---|---|---|---|---|---|---|---|---|
| **Target** | **Arch.** | **Depth.** | $R^2_{\mathrm{MAE}}$ | $R^2_{E^{\mathrm{bot}}_{\mathrm{SOC}}}$ | $R^2_{E^{\mathrm{top}}_{\mathrm{SOC}}}$ | $R^2_{\mathrm{MAE}}$ | $R^2_{E^{\mathrm{bot}}_{\mathrm{SOC}}}$ | $R^2_{E^{\mathrm{top}}_{\mathrm{SOC}}}$ |
| MAE | CNN | 1 | 0.830 | – | – | 0.810 | – | – |
| MAE | CNN | 2 | 0.835 | – | – | 0.813 | – | – |
| MAE | CNN | 3 | 0.835 | – | – | 0.813 | – | – |
| MAE | CNN | 3 | 0.841 | – | – | 0.821 | – | – |
| **MAE** | **CNN** | **4** | **0.843** | – | – | **0.808** | – | – |
| MAE | CNN | 6 | 0.831 | – | – | 0.800 | – | – |
| MAE | ResNet | 2 | 0.831 | – | – | 0.805 | – | – |
| MAE | ResNet | 4 | 0.821 | – | – | 0.800 | – | – |
| MAE | ResNet | 8 | 0.827 | – | – | 0.799 | – | – |
| MAE | DenseNet | 2 | 0.833 | – | – | 0.806 | – | – |
| MAE | DenseNet | 3 | 0.834 | – | – | 0.811 | – | – |
| MAE | DenseNet | 4 | 0.827 | – | – | 0.788 | – | – |
| MAE | VGG | 2 | 0.836 | – | – | 0.804 | – | – |
| MAE | VGG | 3 | 0.829 | – | – | 0.803 | – | – |
| MAE | VGG | 4 | 0.827 | – | – | 0.825 | – | – |
| MAE + $E_{\mathrm{SOC}}$ | CNN | 1 | 0.831 | 0.879 | 0.868 | 0.818 | 0.861 | 0.875 |
| MAE + $E_{\mathrm{SOC}}$ | CNN | 2 | 0.835 | 0.893 | 0.875 | 0.830 | 0.862 | 0.879 |
| MAE + $E_{\mathrm{SOC}}$ | CNN | 3 | 0.835 | 0.893 | 0.875 | 0.830 | 0.862 | 0.879 |
| MAE + $E_{\mathrm{SOC}}$ | **CNN** | **4** | **0.846** | **0.894** | **0.879** | **0.842** | **0.866** | **0.886** |
| MAE + $E_{\mathrm{SOC}}$ | CNN | 6 | 0.841 | 0.889 | 0.879 | 0.836 | 0.865 | 0.881 |
| MAE + $E_{\mathrm{SOC}}$ | ResNet | 2 | 0.838 | 0.900 | 0.880 | 0.824 | 0.868 | 0.874 |
| MAE + $E_{\mathrm{SOC}}$ | ResNet | 4 | 0.828 | 0.892 | 0.884 | 0.825 | 0.848 | 0.871 |
| MAE + $E_{\mathrm{SOC}}$ | ResNet | 8 | 0.819 | 0.902 | 0.887 | 0.817 | 0.870 | 0.869 |
| MAE + $E_{\mathrm{SOC}}$ | DenseNet | 2 | 0.823 | 0.888 | 0.867 | 0.818 | 0.850 | 0.858 |
| MAE + $E_{\mathrm{SOC}}$ | DenseNet | 3 | 0.813 | 0.888 | 0.872 | 0.818 | 0.857 | 0.868 |
| MAE + $E_{\mathrm{SOC}}$ | DenseNet | 4 | 0.824 | 0.889 | 0.877 | 0.825 | 0.866 | 0.882 |
| MAE + $E_{\mathrm{SOC}}$ | VGG | 2 | 0.829 | 0.888 | 0.878 | 0.825 | 0.869 | 0.887 |
| MAE + $E_{\mathrm{SOC}}$ | VGG | 3 | 0.829 | 0.898 | 0.883 | 0.822 | 0.856 | 0.877 |
| MAE + $E_{\mathrm{SOC}}$ | VGG | 4 | 0.820 | 0.891 | 0.889 | 0.813 | 0.855 | 0.874 |
| MAE + $E_{\mathrm{SOC}}$ + $H^d_{\mathrm{SOC}}$ | CNN | 1 | 0.801 | 0.860 | 0.841 | 0.781 | 0.838 | 0.857 |
| MAE + $E_{\mathrm{SOC}}$ + $H^d_{\mathrm{SOC}}$ | CNN | 2 | 0.816 | 0.867 | 0.849 | 0.795 | 0.841 | 0.855 |
| MAE + $E_{\mathrm{SOC}}$ + $H^d_{\mathrm{SOC}}$ | CNN | 3 | 0.816 | 0.867 | 0.849 | 0.795 | 0.841 | 0.855 |
| MAE + $E_{\mathrm{SOC}}$ + $H^d_{\mathrm{SOC}}$ | **CNN** | **4** | **0.817** | **0.877** | **0.860** | **0.804** | **0.858** | **0.865** |
| MAE + $E_{\mathrm{SOC}}$ + $H^d_{\mathrm{SOC}}$ | CNN | 6 | 0.820 | 0.872 | 0.853 | 0.803 | 0.856 | 0.871 |
| MAE + $E_{\mathrm{SOC}}$ + $H^d_{\mathrm{SOC}}$ | ResNet | 2 | 0.817 | 0.874 | 0.851 | 0.801 | 0.845 | 0.854 |
| MAE + $E_{\mathrm{SOC}}$ + $H^d_{\mathrm{SOC}}$ | ResNet | 4 | 0.812 | 0.867 | 0.852 | 0.805 | 0.853 | 0.865 |
| MAE + $E_{\mathrm{SOC}}$ + $H^d_{\mathrm{SOC}}$ | ResNet | 8 | 0.815 | 0.883 | 0.857 | 0.791 | 0.849 | 0.854 |
| MAE + $E_{\mathrm{SOC}}$ + $H^d_{\mathrm{SOC}}$ | DenseNet | 2 | 0.794 | 0.867 | 0.841 | 0.789 | 0.841 | 0.854 |
| MAE + $E_{\mathrm{SOC}}$ + $H^d_{\mathrm{SOC}}$ | DenseNet | 3 | 0.798 | 0.867 | 0.849 | 0.797 | 0.839 | 0.850 |
| MAE + $E_{\mathrm{SOC}}$ + $H^d_{\mathrm{SOC}}$ | DenseNet | 4 | 0.799 | 0.886 | 0.859 | 0.793 | 0.850 | 0.862 |
| MAE + $E_{\mathrm{SOC}}$ + $H^d_{\mathrm{SOC}}$ | VGG | 2 | 0.812 | 0.875 | 0.854 | 0.806 | 0.855 | 0.863 |
| MAE + $E_{\mathrm{SOC}}$ + $H^d_{\mathrm{SOC}}$ | VGG | 3 | 0.806 | 0.879 | 0.855 | 0.810 | 0.852 | 0.876 |
| MAE + $E_{\mathrm{SOC}}$ + $H^d_{\mathrm{SOC}}$ | VGG | 4 | 0.787 | 0.869 | 0.852 | 0.791 | 0.848 | 0.862 |

# S2 Uncertainty in performance metrics

Tables S3–S6 report fold-level uncertainty for the selected BiGRU and 1D-CNN models. All metrics are computed independently on the held-out test fold of each of the $k = 8$ cross-validation splits and are summarised by their mean and standard deviation ($\sigma$) across folds. The 95% confidence interval is obtained from the one-sample $t$-distribution with seven degrees of freedom,

$$\mathrm{CI}_{95} = \bar{x} \pm t_{0.025,\,k-1}\,\frac{\sigma}{\sqrt{k}} = \bar{x} \pm 2.365\,\frac{\sigma}{\sqrt{k}},$$

where $\bar{x}$ is the fold-mean and $k = 8$. $R^2$ values for $E_{\mathrm{SOC}}$ are averages of the per-fold scores for the bottom and top TM atoms; those for $H^d_{\mathrm{SOC}}$ are additionally averaged over the 16 orbital-pair matrix elements. MAErr for $E_{\mathrm{SOC}}$ and $H^d_{\mathrm{SOC}}$ follows the same averaging convention and is expressed in meV. Note that cross-validation folds are not independent—each sample appears in seven training sets—so the intervals are slightly optimistic.

To assess whether the difference in in-domain MAE $R^2$ between the BiGRU and 1D-CNN is statistically meaningful, we apply a paired $t$-test across the eight cross-validation folds. Because both architectures are evaluated on the same held-out test folds, the per-fold scores are paired. In all six target–orbital combinations the difference is not statistically significant at the $\alpha = 0.05$ level ($p$ ranging from 0.08 to 0.85; the closest case is the MAE + $E_{\mathrm{SOC}}$ setting with $s\,p\,d$ orbitals, where the 1D-CNN exceeds the BiGRU by 0.021 with $p = 0.08$). We therefore conclude that the two architectures achieve statistically equivalent in-domain predictive accuracy, consistent with the Rashomon effect, and that the differences discussed in the main text reflect architecture-dependent learned representations rather than differences in overall predictive performance.

**Table S3 | In-domain $R^2$ of the BiGRU models.** Each cell: mean (std) and 95 % CI = mean ± $2.365\,\mathrm{std}/\sqrt{8}$.

| **Target** | **Orb.** | $R^2_{\mathrm{MAE}}$ | $R^2_{E_{\mathrm{SOC}}}$ | $R^2_{H^d_{\mathrm{SOC}}}$ |
|---|---|---|---|---|
| MAE | $s\,p\,d$ | 0.804 (0.055)<br>CI = [0.758, 0.850] | – | – |
| | $d$ | 0.786 (0.051)<br>CI = [0.743, 0.829] | – | – |
| MAE + $E_{\mathrm{SOC}}$ | $s\,p\,d$ | 0.821 (0.051)<br>CI = [0.779, 0.864] | 0.867 (0.045)<br>CI = [0.829, 0.905] | – |
| | $d$ | 0.817 (0.076)<br>CI = [0.754, 0.881] | 0.867 (0.036)<br>CI = [0.837, 0.898] | – |
| MAE + $E_{\mathrm{SOC}}$ + $H^d_{\mathrm{SOC}}$ | $s\,p\,d$ | 0.796 (0.044)<br>CI = [0.759, 0.833] | 0.850 (0.025)<br>CI = [0.829, 0.871] | 0.567 (0.041)<br>CI = [0.532, 0.601] |
| | $d$ | 0.786 (0.050)<br>CI = [0.744, 0.828] | 0.844 (0.034)<br>CI = [0.815, 0.872] | 0.572 (0.054)<br>CI = [0.527, 0.616] |

**Table S4 | In-domain MAErr (meV) of the BiGRU models.** Each cell: mean (std) and 95 % CI. MAErr for $E_{\mathrm{SOC}}$ averaged over the two TM atoms; MAErr for $H^d_{\mathrm{SOC}}$ averaged over the 16 orbital-pair matrix elements.

| **Target** | **Orb.** | $\mathrm{MAErr}_{\mathrm{MAE}}$ | $\mathrm{MAErr}_{E_{\mathrm{SOC}}}$ | $\mathrm{MAErr}_{H^d_{\mathrm{SOC}}}$ |
|---|---|---|---|---|
| MAE | $s\,p\,d$ | 8.84 (1.39)<br>CI = [7.68, 10.00] | – | – |
| | $d$ | 9.26 (0.92)<br>CI = [8.50, 10.03] | – | – |
| $\mathrm{MAE} + E_{\mathrm{SOC}}$ | $s\,p\,d$ | 8.26 (0.86)<br>CI = [7.54, 8.98] | 10.71 (1.23)<br>CI = [9.69, 11.74] | – |
| | $d$ | 8.32 (1.02)<br>CI = [7.46, 9.17] | 10.95 (1.23)<br>CI = [9.93, 11.98] | – |
| $\mathrm{MAE} + E_{\mathrm{SOC}} + H^d_{\mathrm{SOC}}$ | $s\,p\,d$ | 9.14 (0.90)<br>CI = [8.39, 9.89] | 12.09 (1.29)<br>CI = [11.02, 13.17] | 4.070 (0.285)<br>CI = [3.832, 4.309] |
| | $d$ | 9.38 (0.91)<br>CI = [8.62, 10.13] | 12.33 (1.39)<br>CI = [11.17, 13.49] | 4.092 (0.307)<br>CI = [3.835, 4.349] |

**Table S5 | In-domain $R^2$ of the 1D-CNN models.** Each cell: mean (std) and 95 % CI = mean ± $2.365\,\mathrm{std}/\sqrt{8}$.

| **Target** | **Orb.** | $R^2_{\mathrm{MAE}}$ | $R^2_{E_{\mathrm{SOC}}}$ | $R^2_{H^d_{\mathrm{SOC}}}$ |
|---|---|---|---|---|
| MAE | $s\,p\,d$ | 0.808 (0.050)<br>CI = [0.766, 0.850] | – | – |
| | $d$ | 0.801 (0.048)<br>CI = [0.761, 0.842] | – | – |
| $\mathrm{MAE} + E_{\mathrm{SOC}}$ | $s\,p\,d$ | 0.842 (0.029)<br>CI = [0.818, 0.866] | 0.876 (0.046)<br>CI = [0.837, 0.914] | – |
| | $d$ | 0.821 (0.042)<br>CI = [0.786, 0.856] | 0.875 (0.033)<br>CI = [0.847, 0.903] | – |
| $\mathrm{MAE} + E_{\mathrm{SOC}} + H^d_{\mathrm{SOC}}$ | $s\,p\,d$ | 0.804 (0.067)<br>CI = [0.748, 0.860] | 0.862 (0.046)<br>CI = [0.823, 0.901] | 0.601 (0.045)<br>CI = [0.564, 0.639] |
| | $d$ | 0.790 (0.057)<br>CI = [0.743, 0.838] | 0.853 (0.033)<br>CI = [0.825, 0.880] | 0.596 (0.043)<br>CI = [0.560, 0.632] |

**Table S6 | In-domain MAErr (meV) of the 1D-CNN models.** Each cell: mean (std) and 95 % CI. MAErr for $E_{\mathrm{SOC}}$ averaged over the two TM atoms; MAErr for $H^d_{\mathrm{SOC}}$ averaged over the 16 orbital-pair matrix elements.

| **Target** | **Orb.** | $\mathrm{MAErr}_{\mathrm{MAE}}$ | $\mathrm{MAErr}_{E_{\mathrm{SOC}}}$ | $\mathrm{MAErr}_{H^d_{\mathrm{SOC}}}$ |
|---|---|---|---|---|
| MAE | $s\,p\,d$ | 8.38 (0.86)<br>CI = [7.67, 9.10] | – | – |
| | $d$ | 8.69 (0.85)<br>CI = [7.97, 9.40] | – | – |
| $\mathrm{MAE} + E_{\mathrm{SOC}}$ | $s\,p\,d$ | 7.71 (0.83)<br>CI = [7.02, 8.41] | 10.43 (1.15)<br>CI = [9.46, 11.39] | – |
| | $d$ | 8.35 (0.70)<br>CI = [7.77, 8.94] | 10.78 (1.14)<br>CI = [9.82, 11.73] | – |
| $\mathrm{MAE} + E_{\mathrm{SOC}} + H^d_{\mathrm{SOC}}$ | $s\,p\,d$ | 8.57 (1.02)<br>CI = [7.72, 9.43] | 11.35 (1.36)<br>CI = [10.21, 12.49] | 3.853 (0.250)<br>CI = [3.644, 4.062] |
| | $d$ | 9.09 (0.79)<br>CI = [8.43, 9.75] | 11.83 (1.12)<br>CI = [10.90, 12.77] | 3.909 (0.269)<br>CI = [3.685, 4.134] |

## S3 Training Curves

Figure S1 shows the training and validation loss curves for the BiGRU and 1D-CNN models for all eight cross-validation folds and all three target configurations. Both architectures converge smoothly, with the validation loss closely tracking the training loss throughout, indicating no overfitting. Early stopping with a patience of 500 epochs terminates training when the validation loss no longer improves; the stopping epoch, marked by a cross and a vertical dashed line in each panel, varies across folds and target settings but is consistent within each architecture–target combination. The 1D-CNN generally converges in fewer epochs than the BiGRU, reflecting the lower complexity of the convolutional stack relative to the three-layer recurrent encoder.

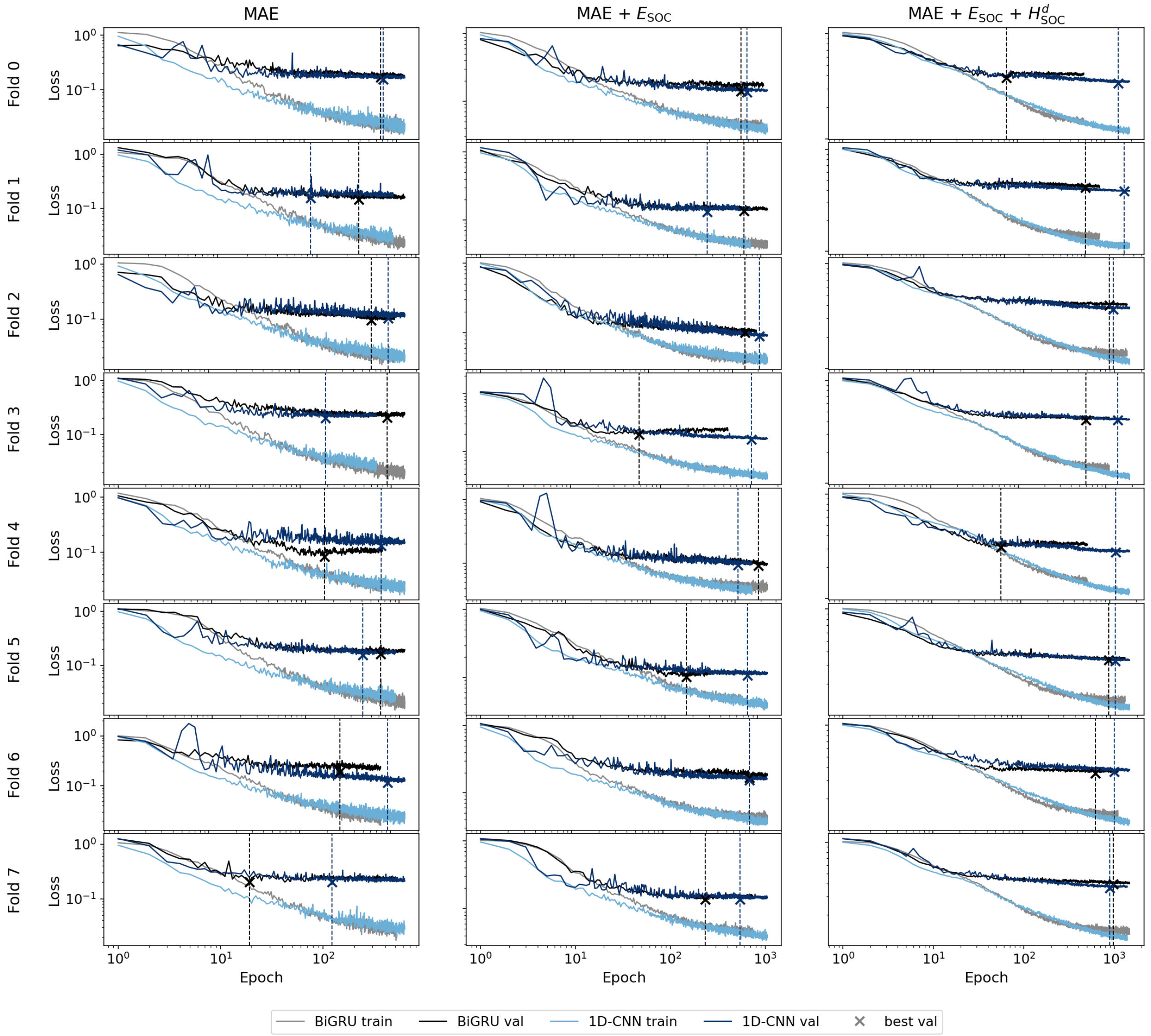


**Figure S1 | Training and validation loss curves for all cross-validation folds.** Each row corresponds to one of the eight folds and each column to one of the three target configurations (MAE, MAE+$E_{SOC}$, and MAE+$E_{SOC}$ + $H^d_{SOC}$). Gray and black curves show the BiGRU training and validation losses, respectively, and light-blue and dark-blue curves show the corresponding 1D-CNN losses. For each model, the cross and vertical dashed line indicate the checkpoint with the lowest validation loss selected by early stopping. Both axes are shown on logarithmic scales.

## S4 Structures

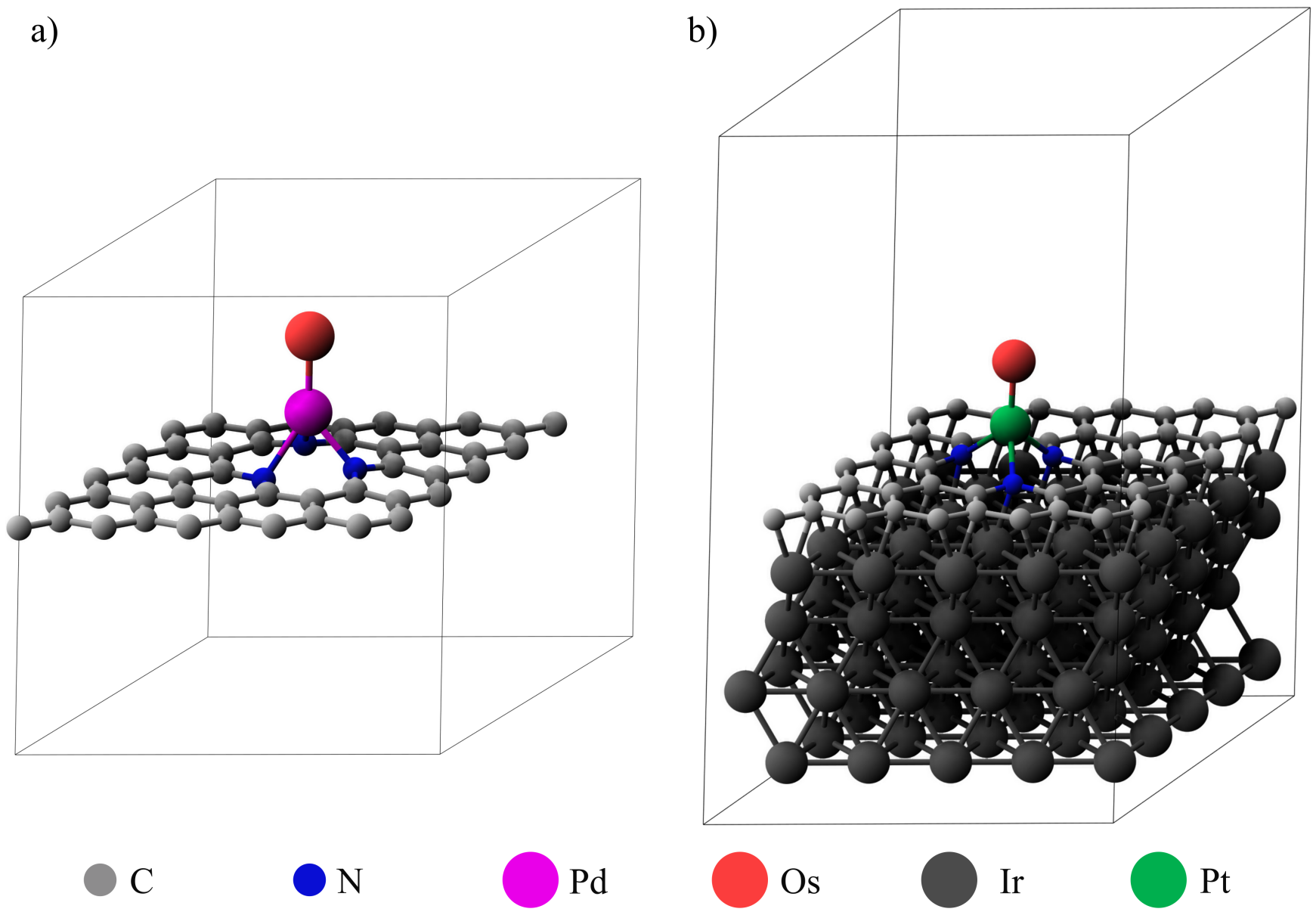


**Figure S2 | Selected supercells rendered using POET.** **a** Freestanding OsPd@NSV, **b** OsPt@NSV@Ir(111).

# S5 Energy-resolved SHAP

### SHAP decomposition

SHapley Additive exPlanations (SHAP) [1, 2] decomposes a model prediction into additive feature contributions relative to a background expectation. For a model $f$ receiving the Fermi-reordered multichannel SR-DOS input $\mathbf{X} \in \mathbb{R}^{T \times C}$, the prediction is expressed as

$$\hat{y} = \mathbb{E}_{\mathbf{X}' \sim \mathcal{B}}\big[f(\mathbf{X}')\big] + \sum_{t=1}^{T} \sum_{c=1}^{C} \phi_{t,c}, \tag{S1}$$

where $T = 100$ is the number of positions along the reordered distance-from-$E_{\mathrm{F}}$ axis, $C$ is the number of input channels (including contributions from both TM atoms), $\phi_{t,c}$ is the SHAP value assigned to the DOS intensity at energy index $t$ and channel $c$, and $\mathcal{B}$ is the background dataset. A positive $\phi_{t,c}$ indicates that the corresponding spectral weight at $(t, c)$ shifts the prediction toward a higher value relative to the background expectation, whereas a negative value indicates a suppressive contribution. Because contributions of opposite sign can cancel upon summation, the absolute SHAP magnitude $|\phi_{t,c}|$ is used to quantify attributed importance independently of the direction of the contribution.

### Background distribution

The background dataset $\mathcal{B}$ consists of weak-anisotropy systems with MAE $\in [-25, 25]$ meV. The resulting SHAP values describe feature contributions relative to the model response expected for this reference population. For example, a positive $\phi_{t,c}$ assigned to a given energy and orbital channel indicates that the corresponding spectral feature shifts the predicted MAE toward more positive values relative to the background expectation. Attribution patterns are analyzed separately for high-positive-MAE systems (MAE $> 25$ meV) and strongly negative-MAE systems (MAE $< -25$ meV).

### Gradient-based SHAP

SHAP values are computed using the GradientExplainer algorithm [2], which estimates feature attributions for differentiable models using expected gradients. The analysis is performed using models retrained on the complete implicit-substrate dataset, as described in the *Training procedure* section. Attributions are initially obtained in the standardized target space used during training and are converted to physical units by multiplying them by the standard deviation of the corresponding target scaler. For MAE, the resulting SHAP values are reported in meV.

## S6 DOS–SHAP overlays

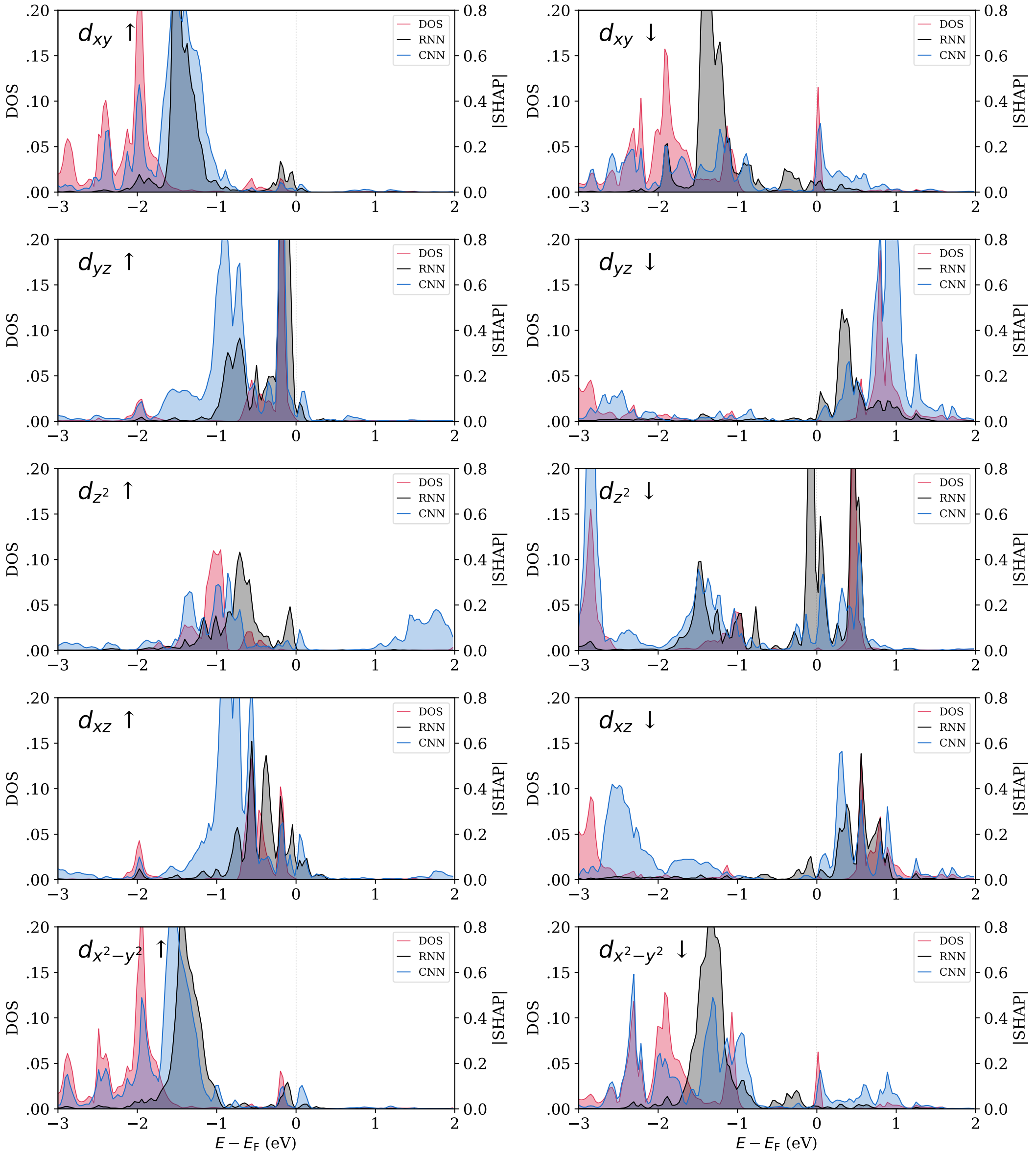


**Figure S3 | Energy-resolved DOS–SHAP overlays.** The profiles correspond to the bottom TM atom of the high-MAE OsPd@NSV dimer with a DFT-computed MAE of 133.0 meV. Columns and rows distinguish the spin channels and individual $d$ orbitals, respectively.

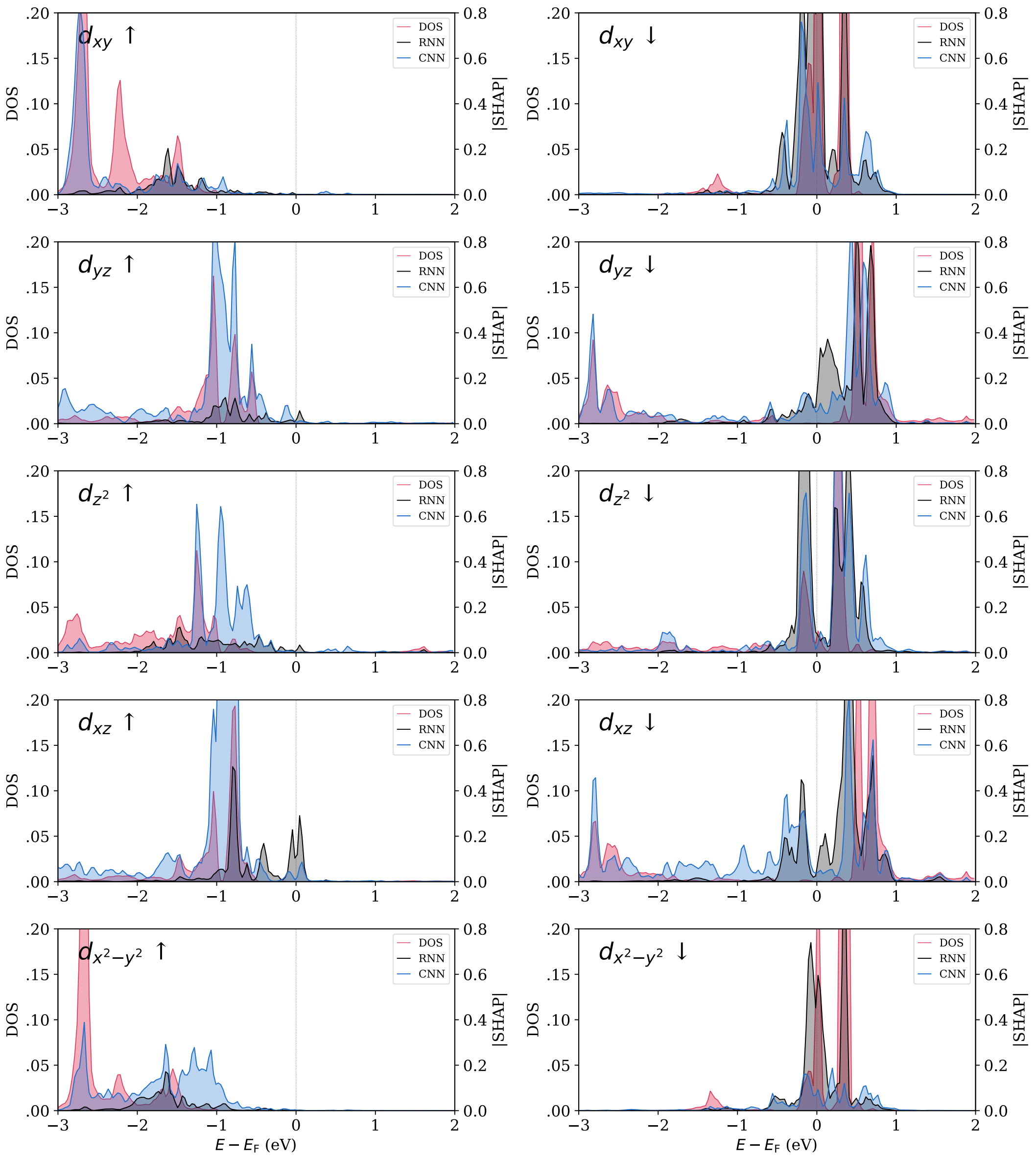


**Figure S4 | Energy-resolved DOS–SHAP overlays.** The profiles correspond to the top TM atom of the strongly negative-MAE OsPt@NSV dimer with a DFT-computed MAE of −75.7 meV. Rows and columns distinguish the spin channels and individual $d$ orbitals, respectively.

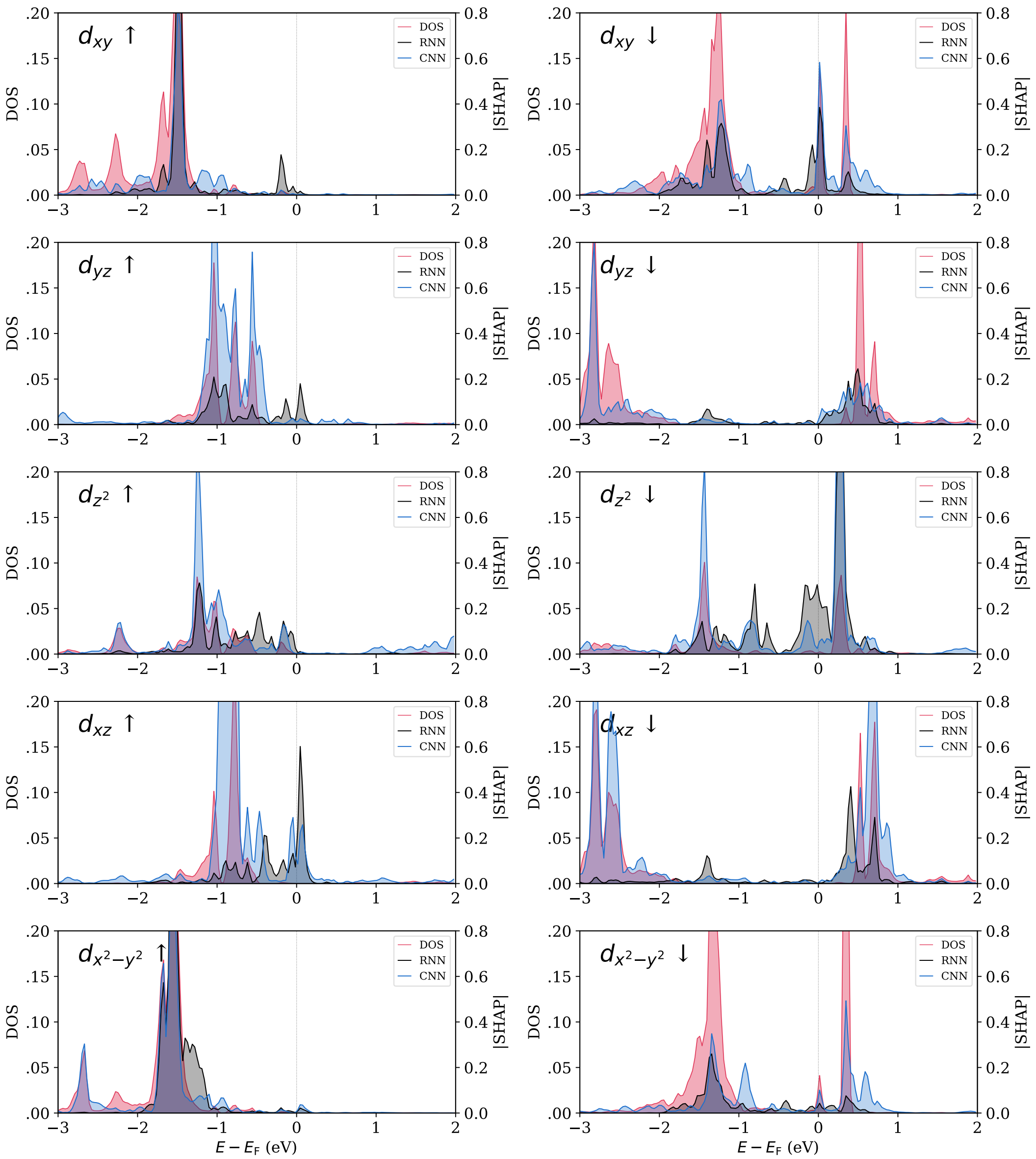


**Figure S5 | Energy-resolved DOS–SHAP overlays.** The profiles correspond to the bottom TM atom of the strongly negative-MAE OsPt@NSV dimer with a DFT-computed MAE of −75.7 meV. Columns and rows distinguish the spin channels and individual *d* orbitals, respectively.

# S7 Additional DOS–SHAP overlays

Energy-resolved DOS–SHAP overlays are presented for additional systems including high-positive-MAE systems, systems near the lower boundary of the high-positive-MAE regime, and strongly negative-MAE systems. Each figure resolves the overlays by TM atom, spin channel, and $d$ orbital.

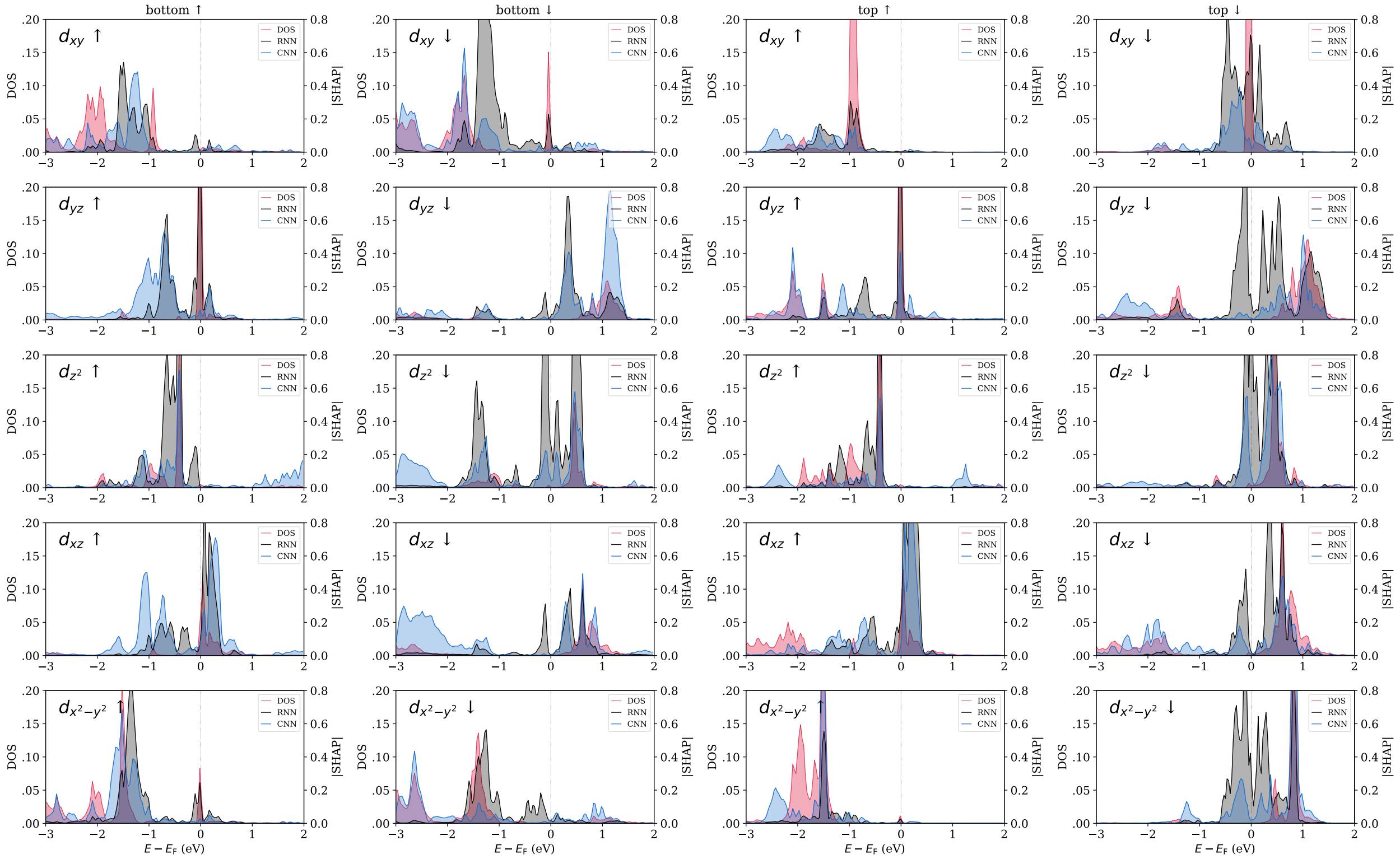


**Figure S6 | Energy-resolved DOS–SHAP overlays.** Results are shown for the bottom and top TM atoms of the highest-MAE OsPt@NSV system with a DFT-computed MAE of 158.2 meV. Rows distinguish the individual $d$ orbitals. The columns correspond to the spin-up and spin-down channels of the bottom TM atom, followed by the corresponding channels of the top TM atom.

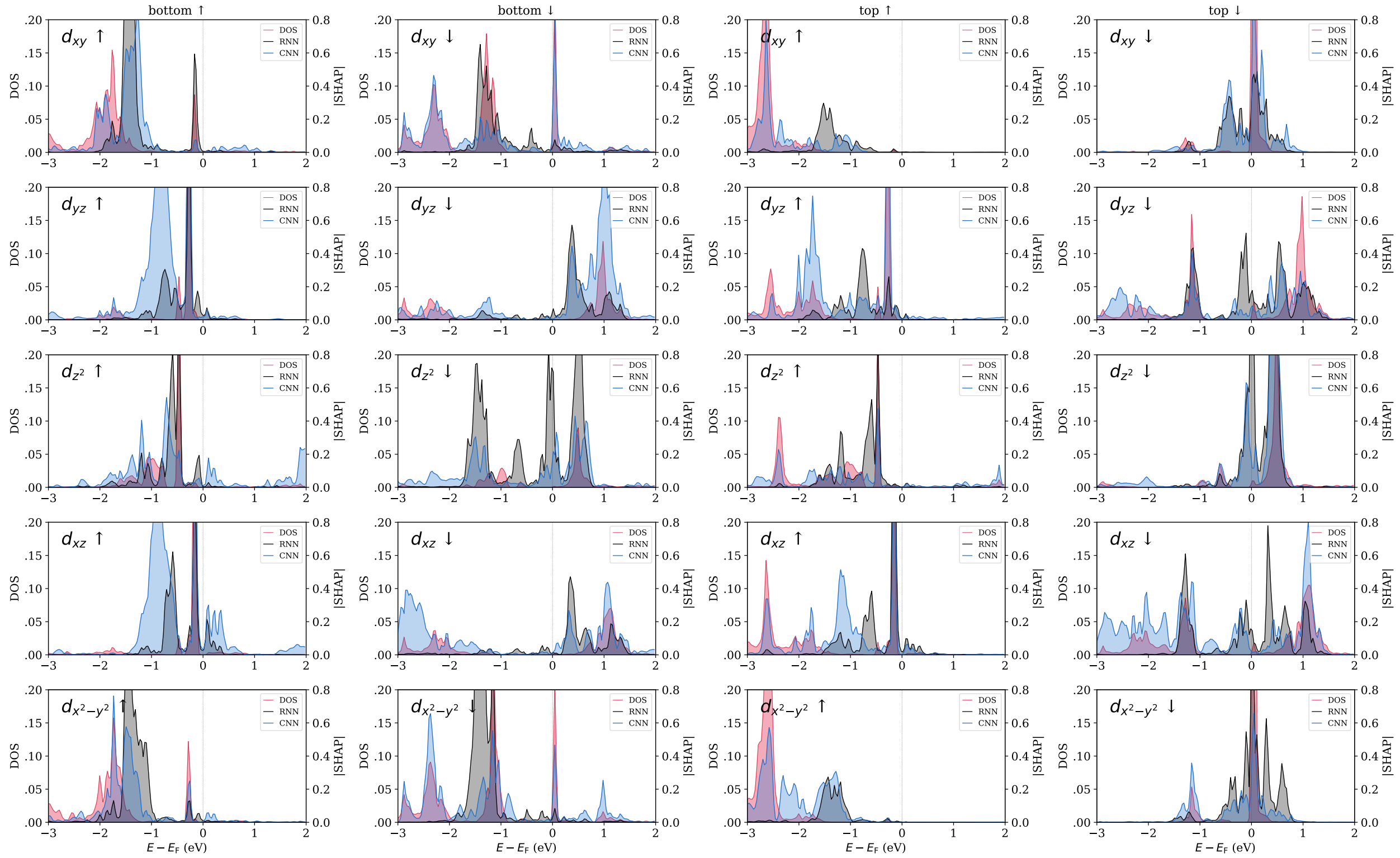

**Figure S7 | Energy-resolved DOS–SHAP overlays.** Results are shown for the bottom and top TM atoms of a high-MAE OsPt@NSV dimer with a DFT-computed MAE of 126.6 meV. Rows distinguish the individual $d$ orbitals. The columns correspond to the spin-up and spin-down channels of the bottom TM atom, followed by the corresponding channels of the top TM atom.

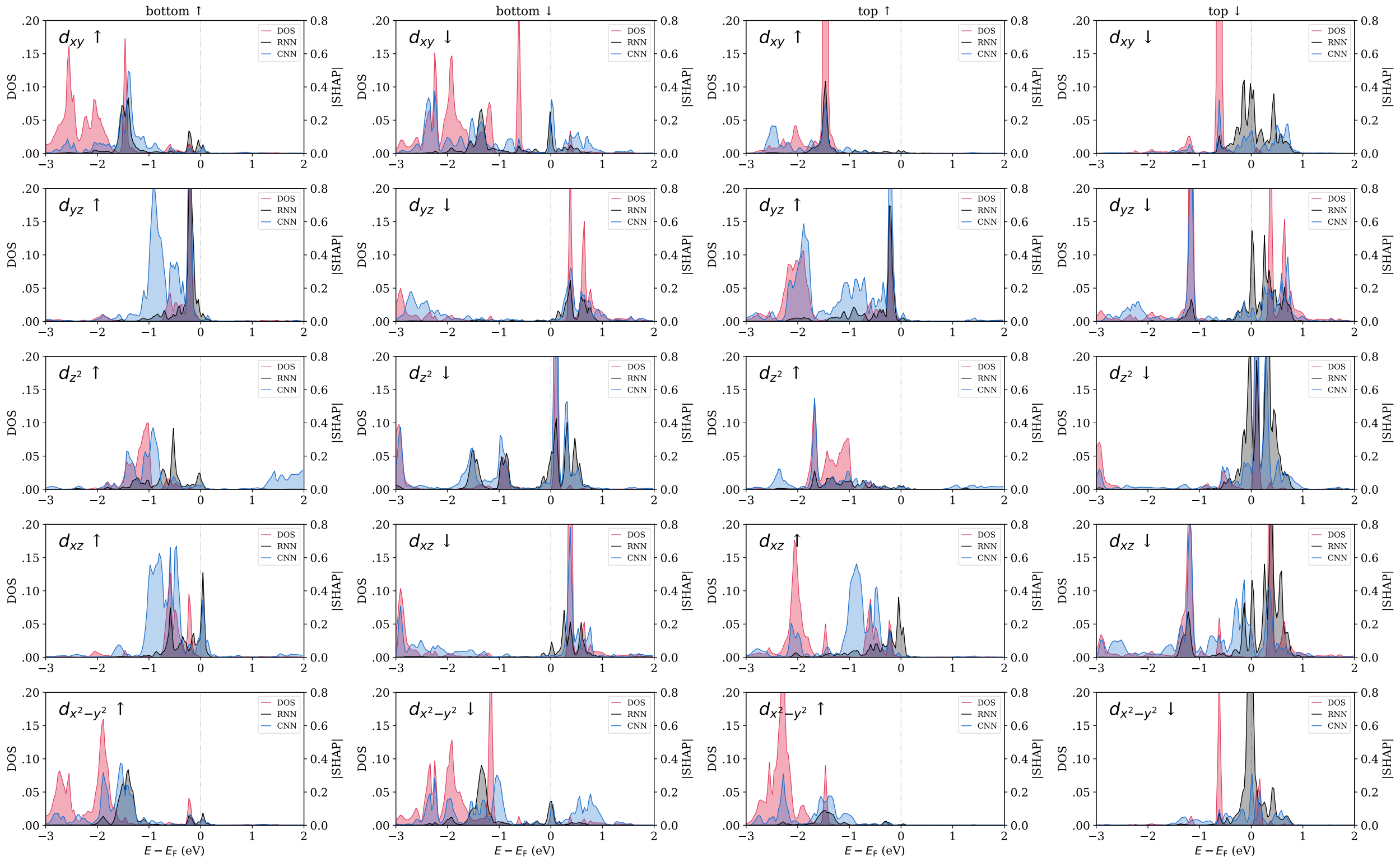


**Figure S8 | Energy-resolved DOS–SHAP overlays.** Results are shown for the bottom and top TM atoms of an IrPd@NSV dimer near the lower boundary of the high-positive-MAE regime, with a DFT-computed MAE of 27.4 meV. Rows distinguish the individual $d$ orbitals. The columns correspond to the spin-up and spin-down channels of the bottom TM atom, followed by the corresponding channels of the top TM atom.

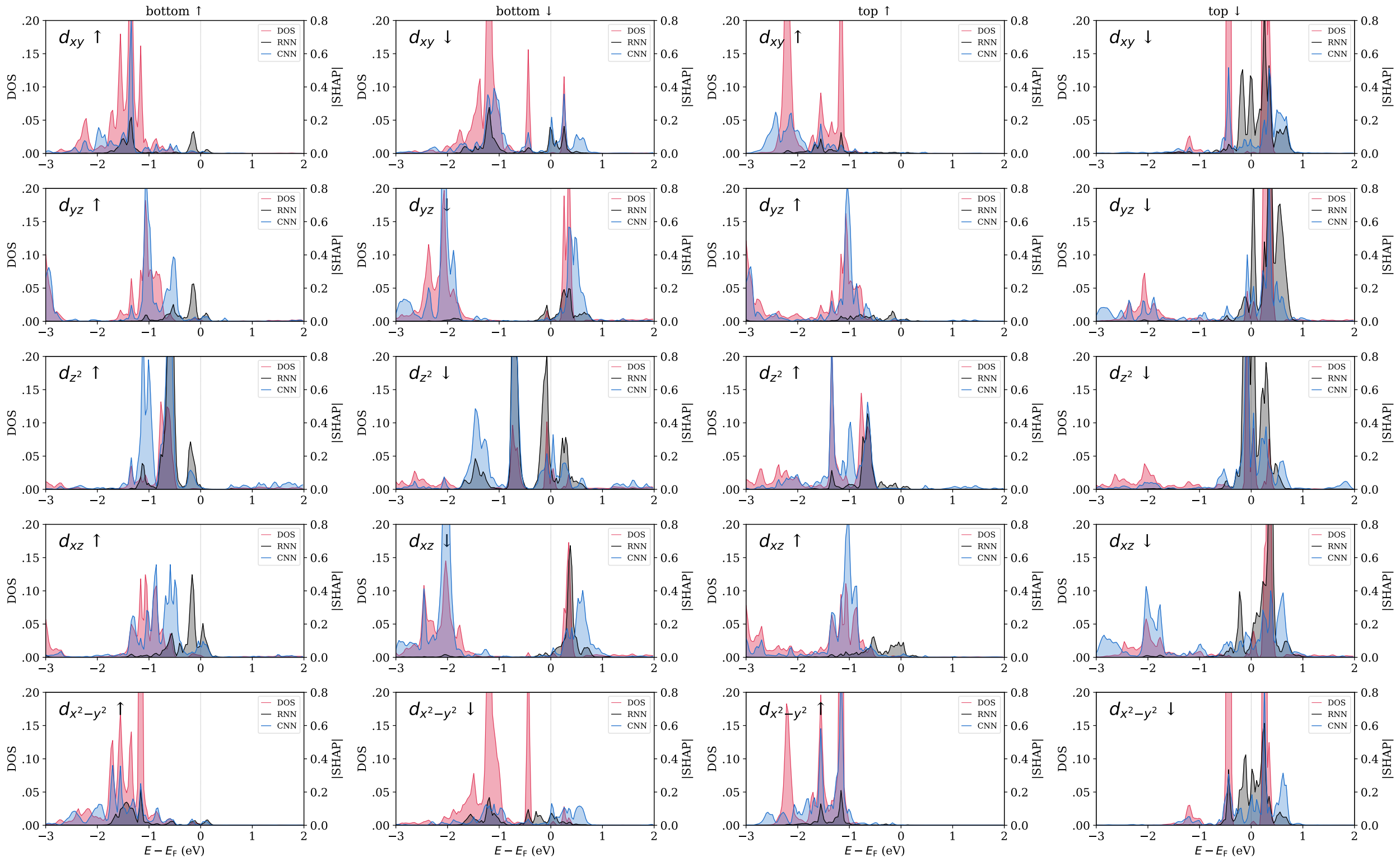


**Figure S9 | Energy-resolved DOS–SHAP overlays.** Results are shown for the bottom and top TM atoms of an IrPd@NSV dimer near the lower boundary of the high-positive-MAE regime, with a DFT-computed MAE of 28.9 meV. Rows distinguish the individual $d$ orbitals. The columns correspond to the spin-up and spin-down channels of the bottom TM atom, followed by the corresponding channels of the top TM atom.

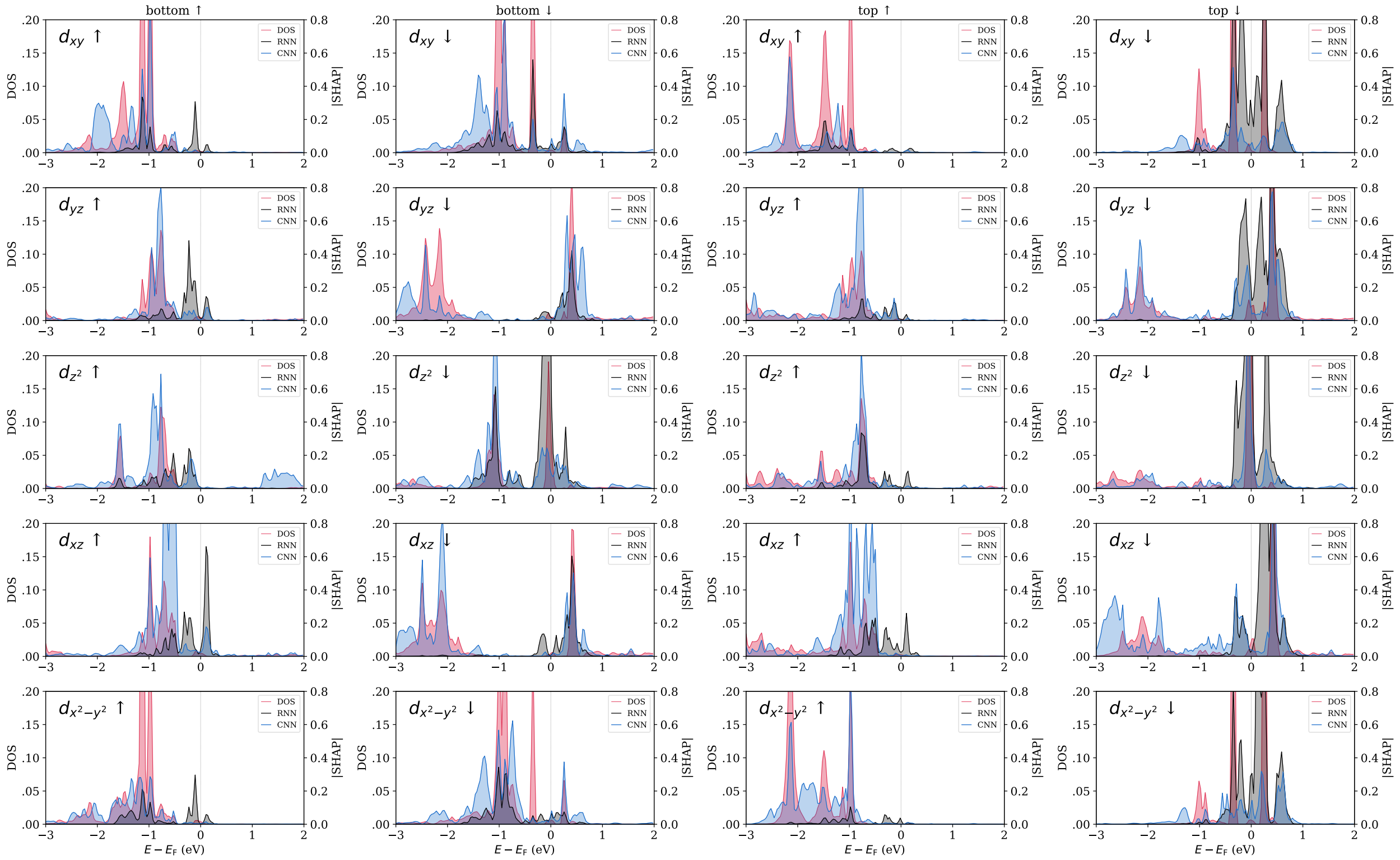


**Figure S10 | Energy-resolved DOS–SHAP overlays.** Results are shown for the bottom and top TM atoms of a strongly negative-MAE IrPt@NSV dimer with a DFT-computed MAE of −56.2 meV. Rows distinguish the individual $d$ orbitals. The columns correspond to the spin-up and spin-down channels of the bottom TM atom, followed by the corresponding channels of the top TM atom.

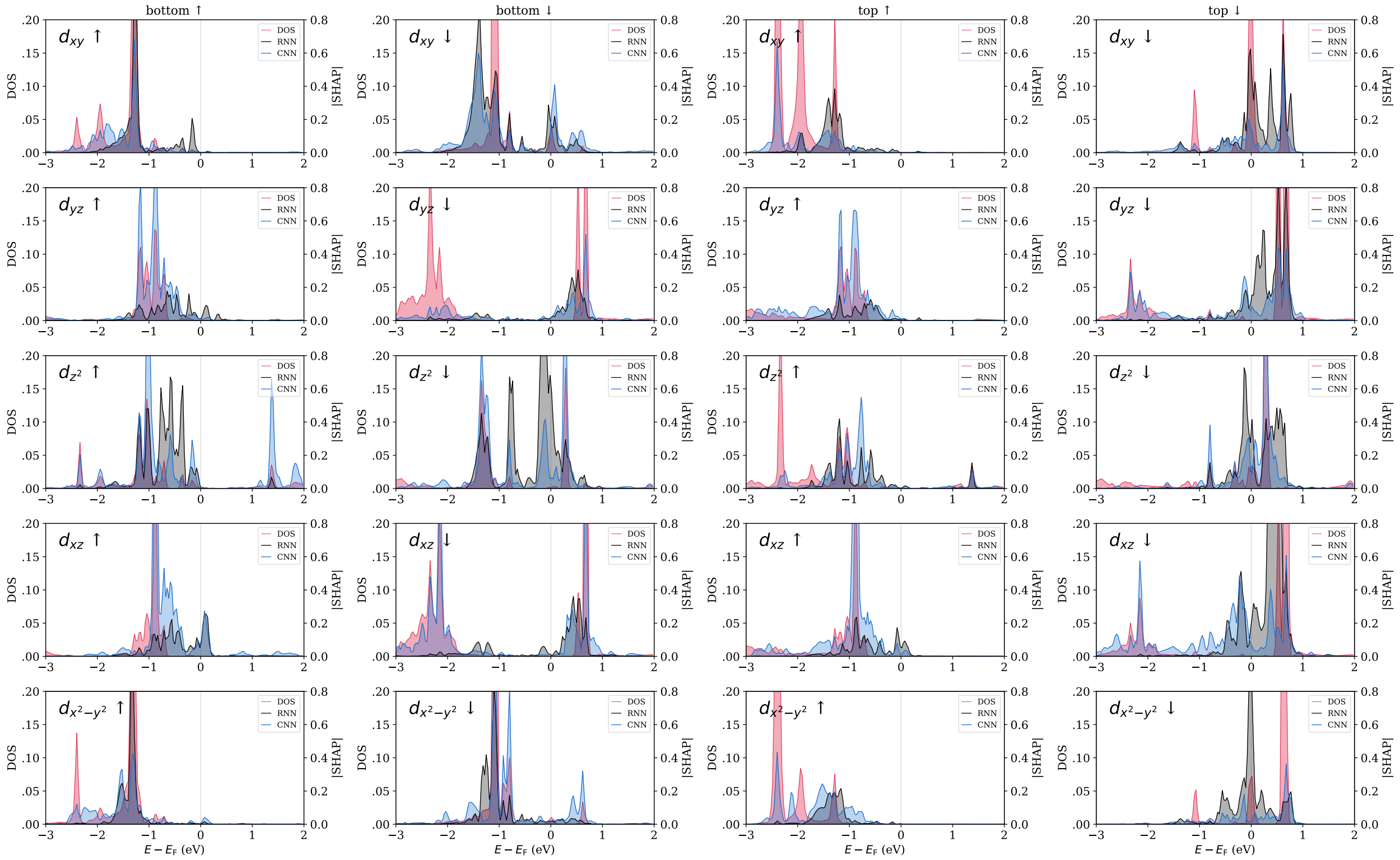


**Figure S11 | Energy-resolved DOS–SHAP overlays.** Results are shown for the bottom and top TM atoms of a strongly negative-MAE OsPt@NSV dimer with a DFT-computed MAE of −64.8 meV. Rows distinguish the individual $d$ orbitals. The columns correspond to the spin-up and spin-down channels of the bottom TM atom, followed by the corresponding channels of the top TM atom.

# S8 SHAP and MFT comparison

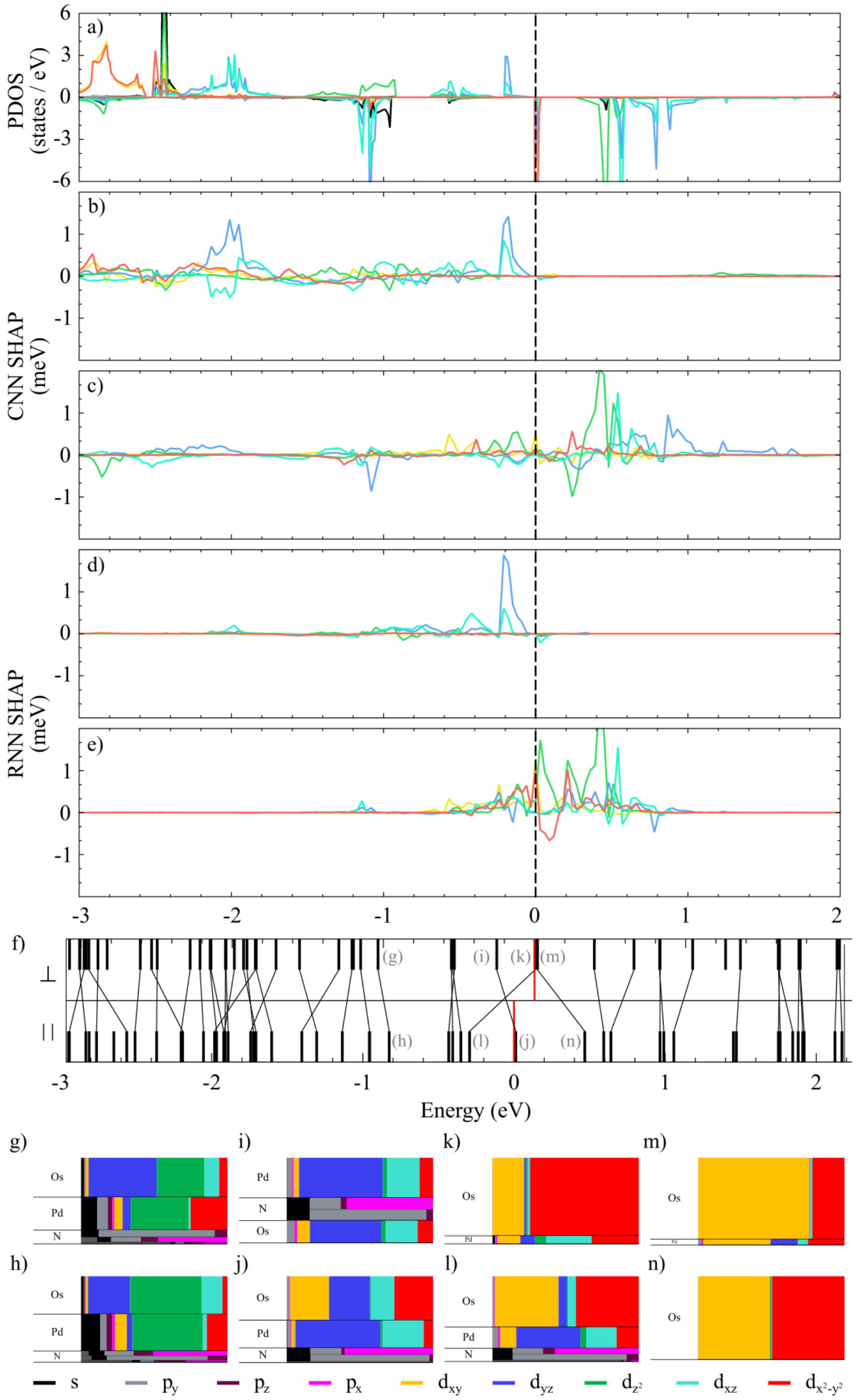


**Figure S12 | Orbital-resolved comparison of ML attributions and DFT-computed contributions to the magnetic anisotropy energy (MAE) of the Os atom in OsPd@NSV.** The DFT-computed MAE of the system is 133.0 meV. **a**, Spin- and orbital-resolved scalar-relativistic density of states (SR-DOS). **b**, **c**, SHAP attributions obtained from the 1D-CNN for the spin-up and spin-down channels, respectively. **d**, **e**, Corresponding SHAP attributions obtained from the BiGRU. **f**, Eigenvalue spectra extracted from magnetic-force-theorem (MFT) [3, 4] calculations using POET [5] for magnetization perpendicular (top) and parallel (bottom) to the TM dimer bond. Each spectrum is referenced to its corresponding Fermi level ($E_F$; red lines), and the energy axes are aligned to account for the difference in $E_F$. Labels (g)–(n) in **f** identify states with relevant contributions to the MAE. Their atom- and orbital-resolved decompositions are shown in **(g–n)**, with box dimensions proportional to the corresponding orbital weights.

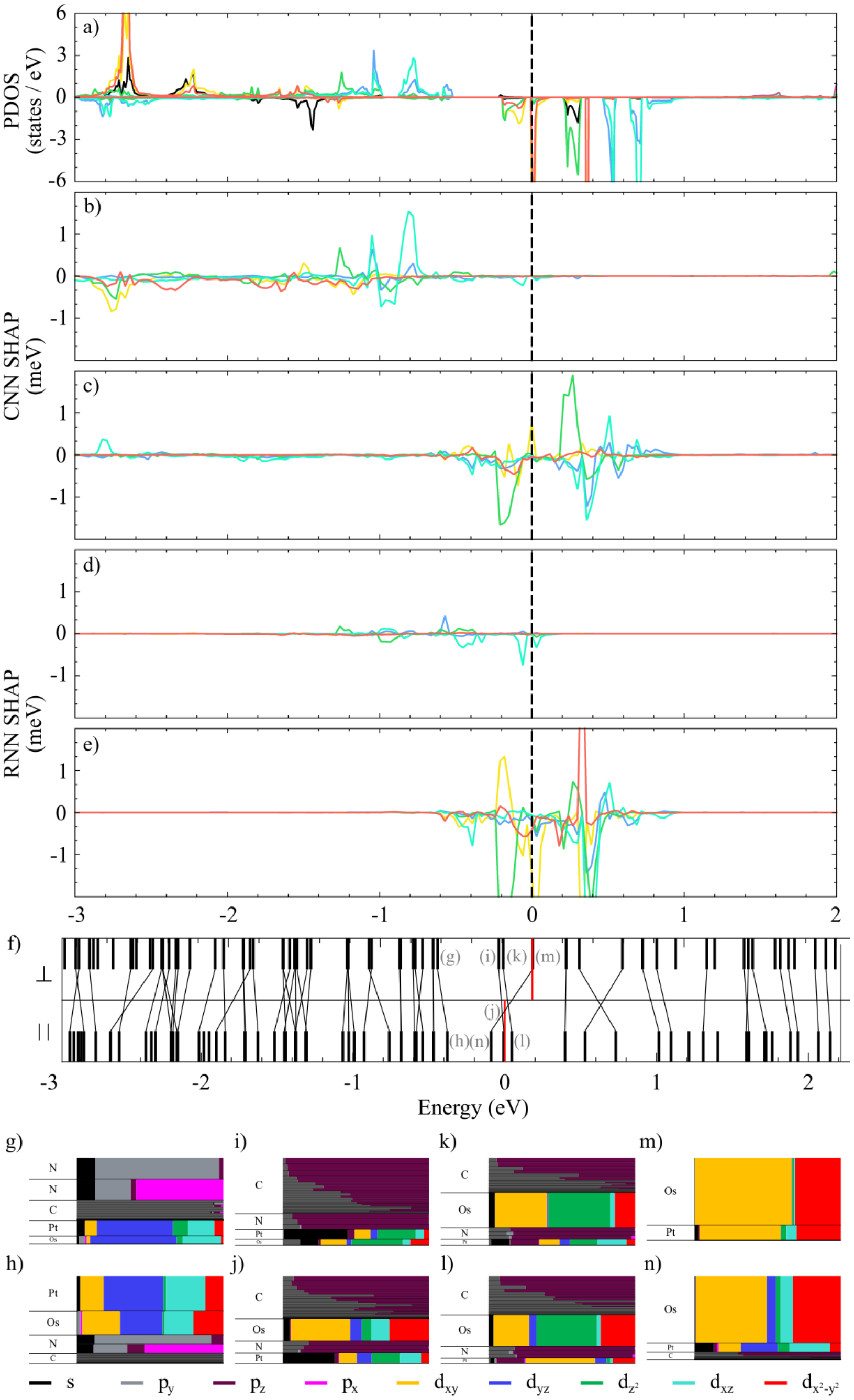


**Figure S13 | Orbital-resolved comparison of ML attributions and DFT-computed contributions to the magnetic anisotropy energy (MAE) of the Os atom in OsPt@NSV.** The DFT-computed MAE of the system is −75.7 meV. **a**, Spin- and orbital-resolved scalar-relativistic density of states (SR-DOS). **b**, **c**, SHAP attributions obtained from the 1D-CNN for the spin-up and spin-down channels, respectively. **d**, **e**, Corresponding SHAP attributions obtained from the BiGRU. **f**, Eigenvalue spectra extracted from magnetic-force-theorem (MFT) [3, 4] calculations performed using POET [5] for magnetization perpendicular (top) and parallel (bottom) to the TM–TM bond. Each spectrum is referenced to its corresponding Fermi level ($E_{\mathrm{F}}$; red lines), and the energy axes are aligned to account for the difference between the two Fermi levels. Labels (g)–(n) in **f** identify states with relevant contributions to the MAE. Their atom- and orbital-resolved decompositions are shown in **g**–**n**, with box areas proportional to the corresponding orbital weights.

## S9 Controlled DOS perturbations

**Table S7 | Parameters of the controlled DOS perturbations.** The listed parameters define the Gaussian modifications introduced in Eq. (1) of the main text and correspond to the modified DOS profiles shown in the indicated figures. $L$ and $W$ denote the center and width of the perturbation, respectively, and are given in eV. $H$ is the limiting signed amplitude in the units of the DOS. Positive and negative values of $H$ correspond to the addition and attenuation of spectral weight, respectively.

| Figure | $L$ | $W$ | $H$ |
|---|---|---|---|
| Fig. 5a | −0.20 | 0.20 | −0.30 |
| Fig. 5c | −0.10 | 0.05 | 0.30 |
| Fig. S14a | −2.00 | 0.25 | −0.17 |
| Fig. S14c | 1.00 | 0.15 | 0.30 |
| Fig. S15a | −0.16 | 0.20 | −0.09 |
| Fig. S15c | 0.40 | 0.05 | 0.50 |
| Fig. S16a | −1.45 | 0.10 | −0.10 |
| Fig. S16c | 1.00 | 0.10 | 0.30 |
| Fig. S18a | −2.00 | 0.25 | −0.17 |
| Fig. S18c | 1.00 | 0.15 | 0.30 |
| Fig. S19a | −0.16 | 0.20 | −0.09 |
| Fig. S19c | 0.40 | 0.05 | 0.50 |
| Fig. S20a | −1.45 | 0.10 | −0.10 |
| Fig. S20c | 1.00 | 0.10 | 0.30 |

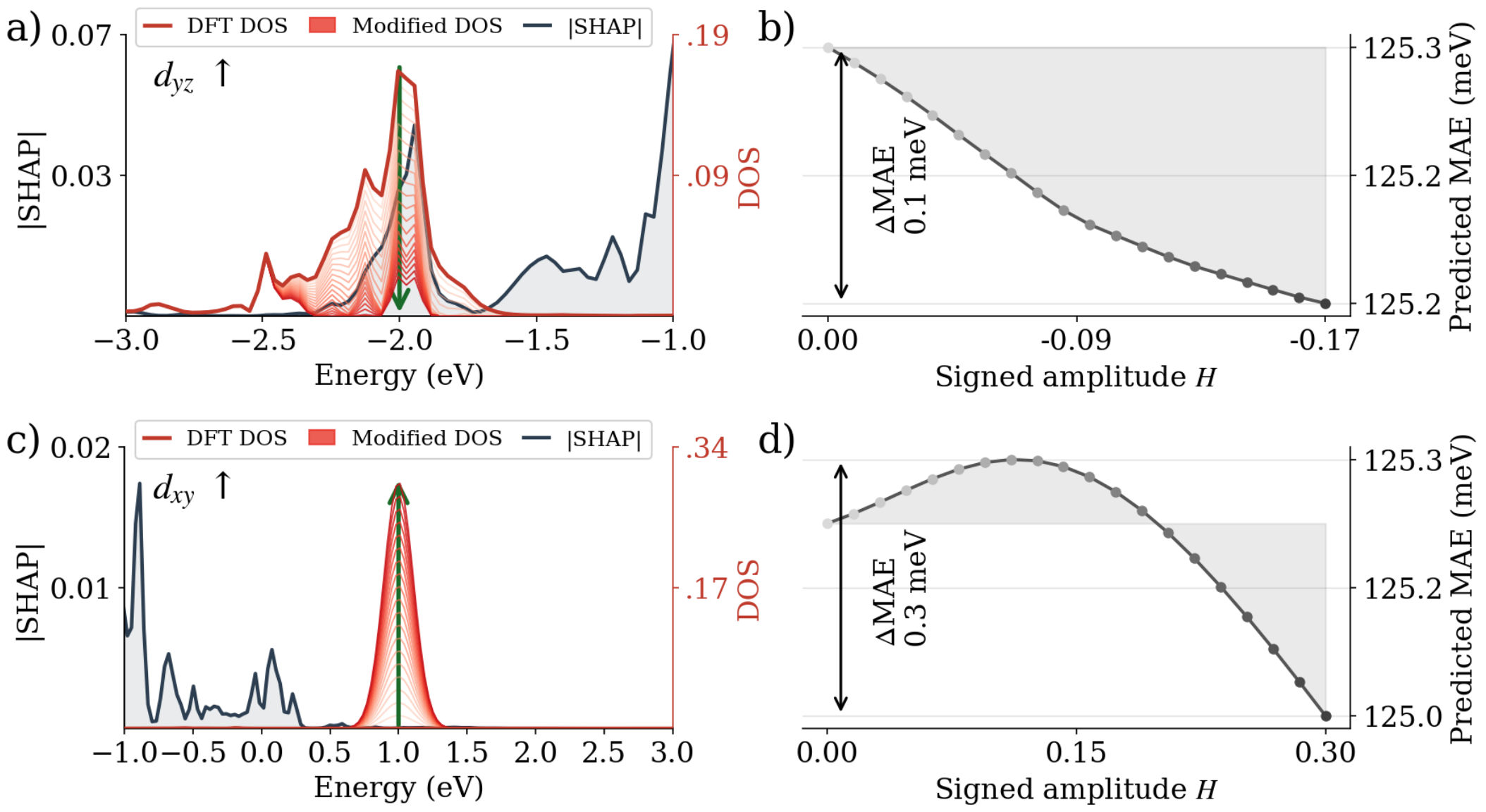


**Figure S14 | Controlled DOS perturbations.** Results are shown for the top TM atom of the high-MAE OsPd@NSV dimer, with a DFT-computed MAE of 133.0 meV, evaluated using the BiGRU. **a**, Attenuation of an existing DOS peak; **c**, introduction of an artificial DOS peak. Both modifications target regions with negligible SHAP attribution. **b**, **d**, Corresponding MAE predictions as functions of the perturbation amplitude $H$.

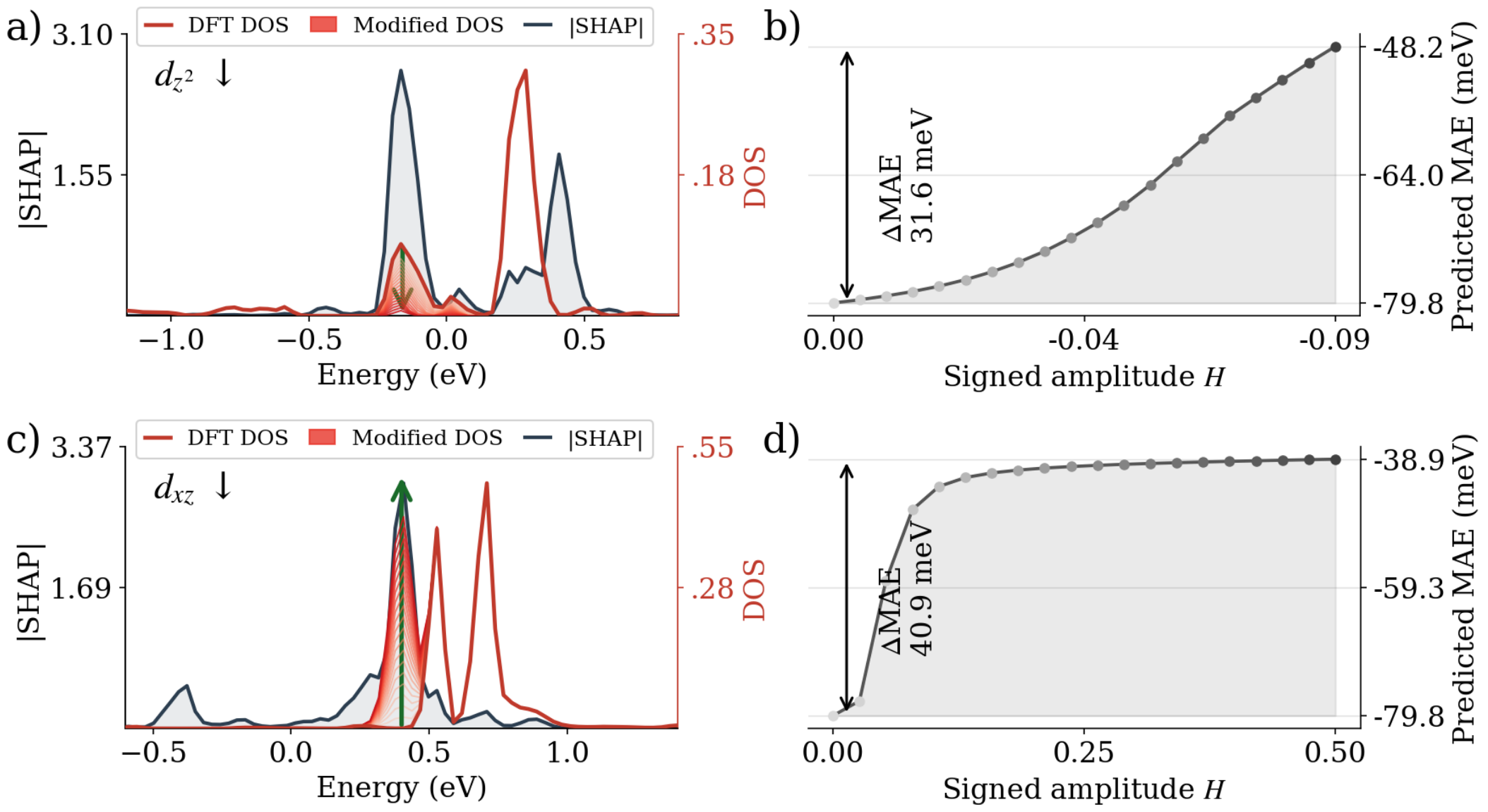


**Figure S15 | Controlled DOS perturbations.** Results are shown for the top TM atom of the strongly negative-MAE OsPt@NSV dimer, with a DFT-computed MAE of −75.7 meV, evaluated using the BiGRU. **a**, Attenuation of an occupied DOS feature in a region with strong SHAP attribution; **c**, introduction of spectral weight into another SHAP-active energy window. **b**, **d**, Corresponding MAE predictions as functions of the perturbation amplitude $H$.

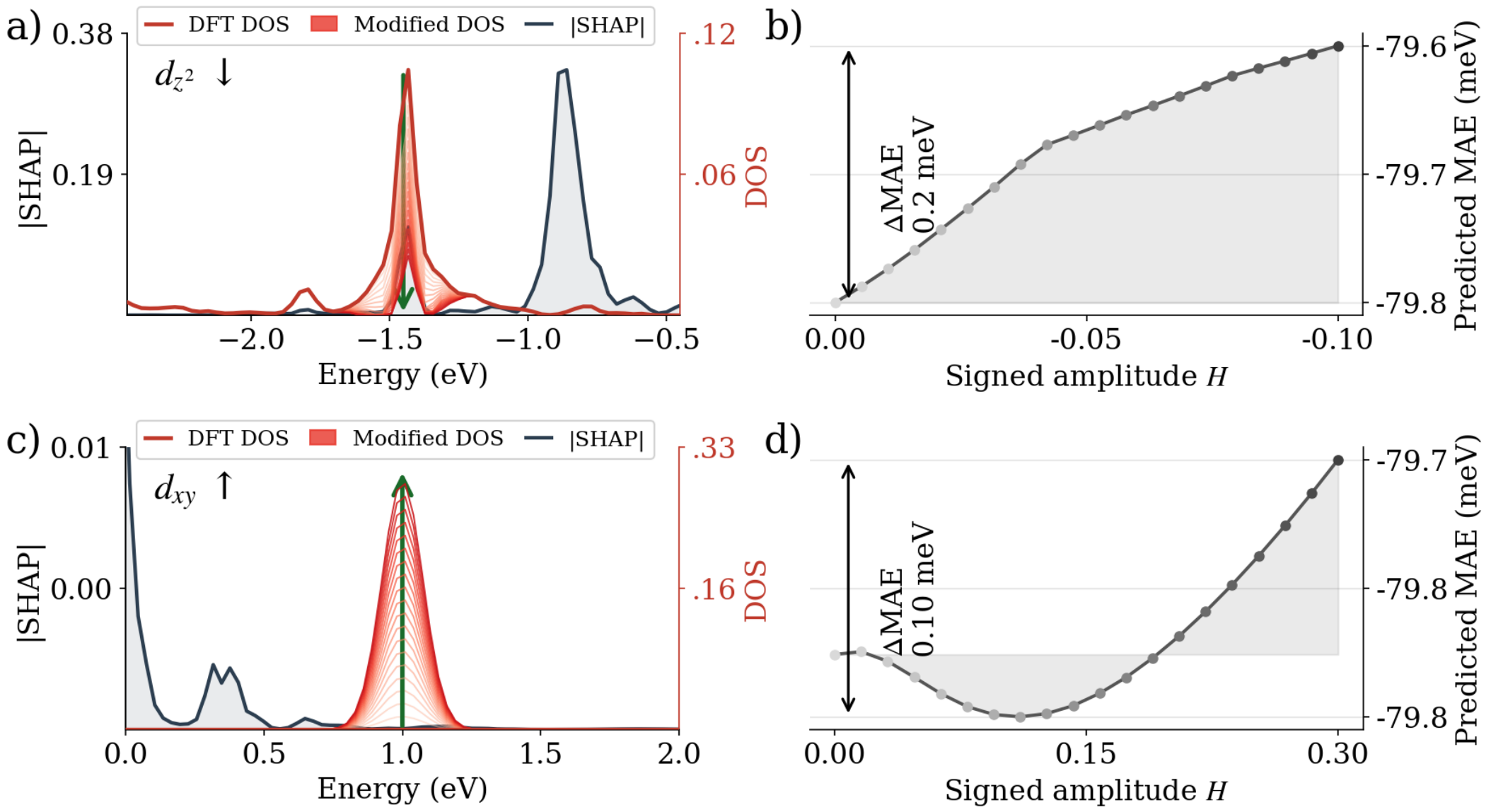


**Figure S16 | Controlled DOS perturbations in SHAP-inactive regions.** Results are shown for the strongly negative-MAE OsPt@NSV dimer, with a DFT-computed MAE of −75.7 meV, evaluated using the BiGRU. **a**, Attenuation of a DOS peak for the bottom TM atom; **c**, introduction of an artificial DOS peak for the top TM atom. **b**, **d**, Corresponding MAE predictions as functions of the perturbation amplitude $H$.

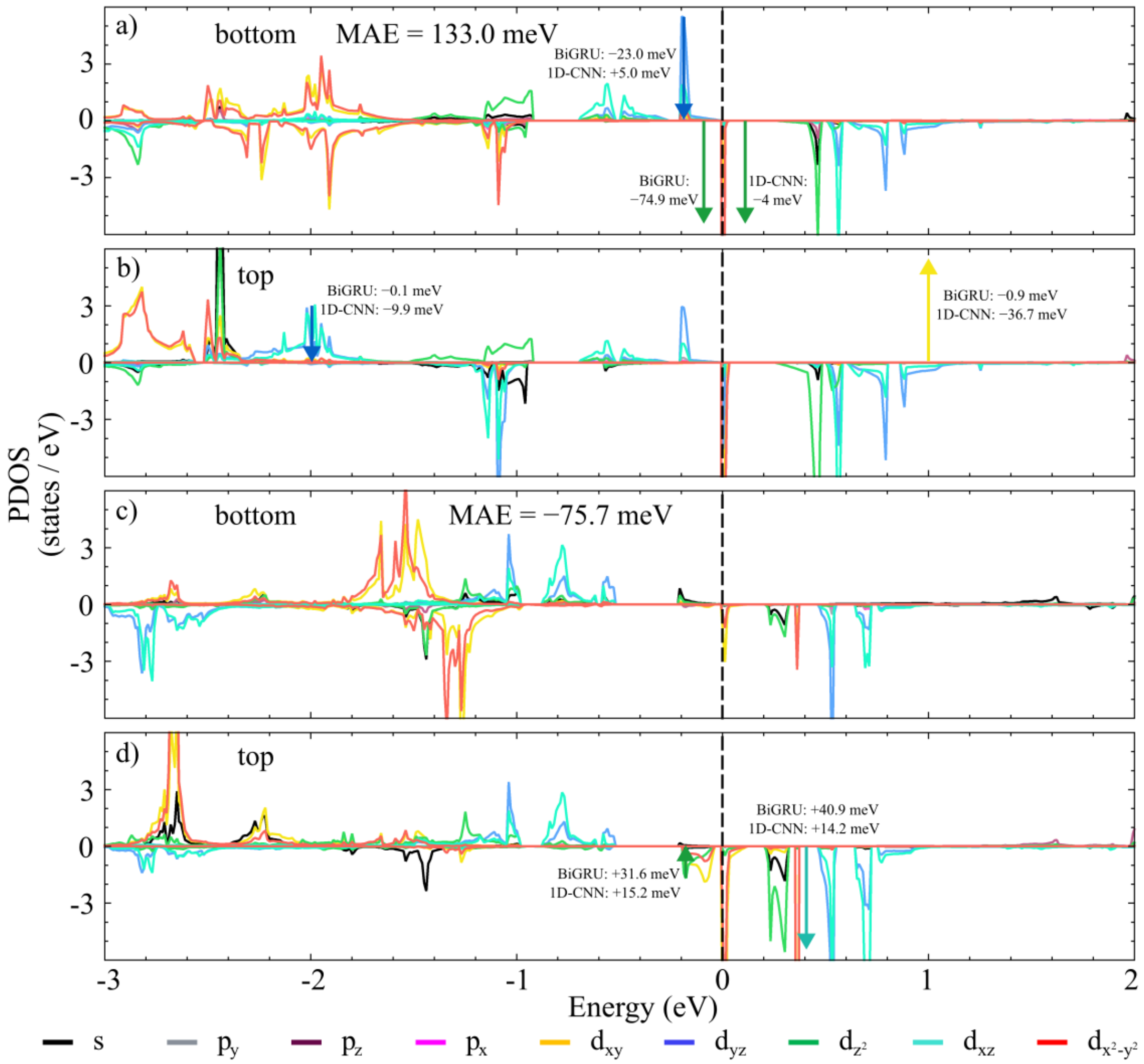


**Figure S17 | Orbital-resolved SR-DOS and selected spectral perturbations.** **a**, **b**, Orbital-resolved SR-DOS of the bottom and top TM atoms, respectively, in the high-MAE OsPd@NSV dimer with a DFT-computed MAE of 133.0 meV. **c**, **d**, Corresponding SR-DOS of the bottom and top TM atoms in OsPt@NSV with a DFT-computed MAE of −75.7 meV. Positive and negative DOS values represent the two spin channels. The vertical dashed line marks the Fermi level, $E_{\mathrm{F}}$. Colored arrows indicate the orbital channels and energy positions selected for attenuation of existing DOS features or introduction of spectral weight into initially depleted regions. Labels report the resulting changes in the MAE predicted by the BiGRU and 1D-CNN at the maximum perturbation amplitude. Orbital contributions are identified by the color legend below the panels.

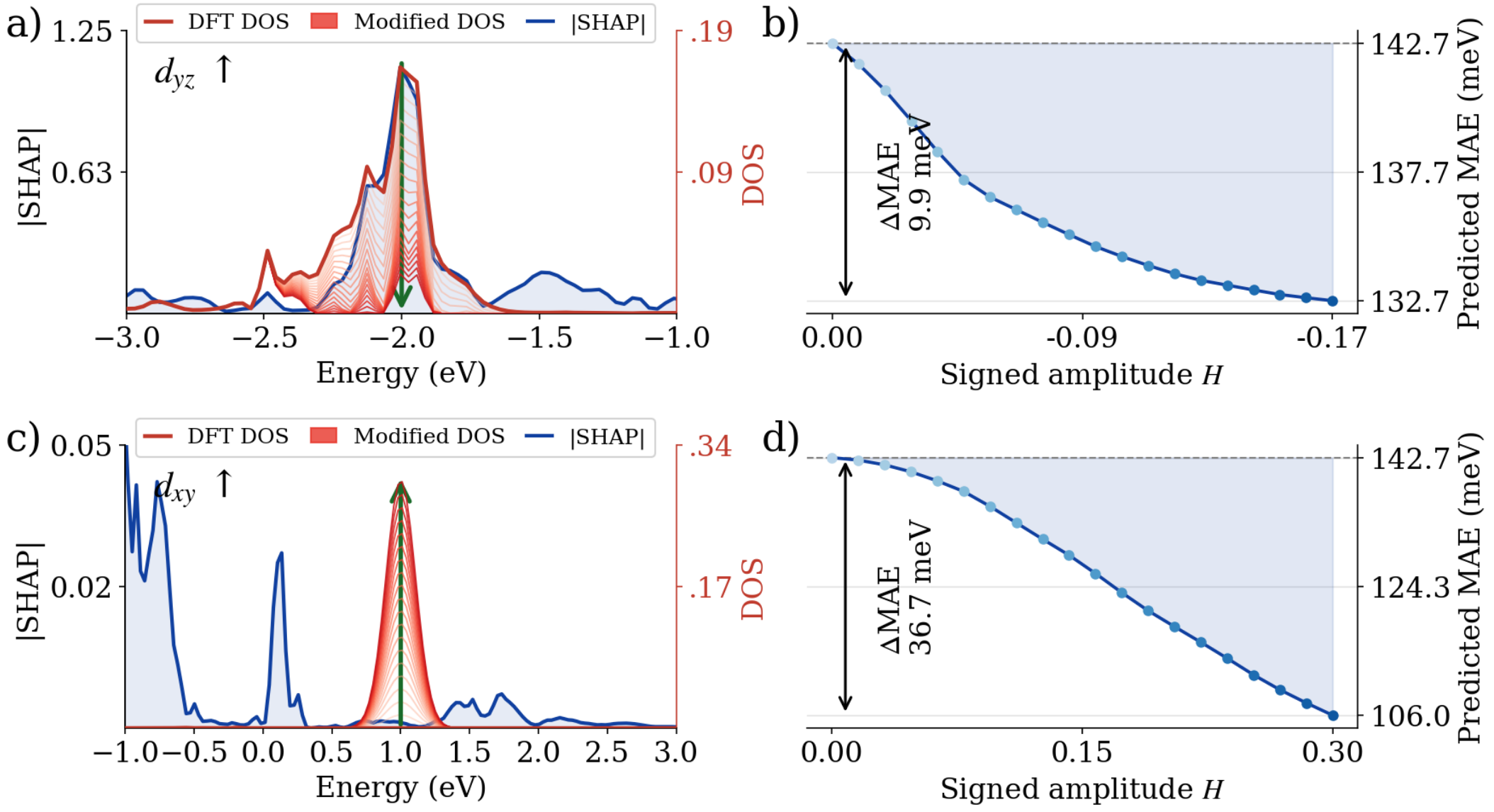


**Figure S18 | Controlled DOS perturbations.** Results are shown for the top TM atom of the high-MAE OsPd@NSV dimer, with a DFT-computed MAE of 133.0 meV, evaluated using the 1D-CNN. **a**, Attenuation of a DOS peak carrying substantial SHAP attribution; **c**, introduction of spectral weight into an initially depleted interval with negligible attribution. **b**, **d**, Corresponding MAE predictions as functions of the perturbation amplitude $H$.

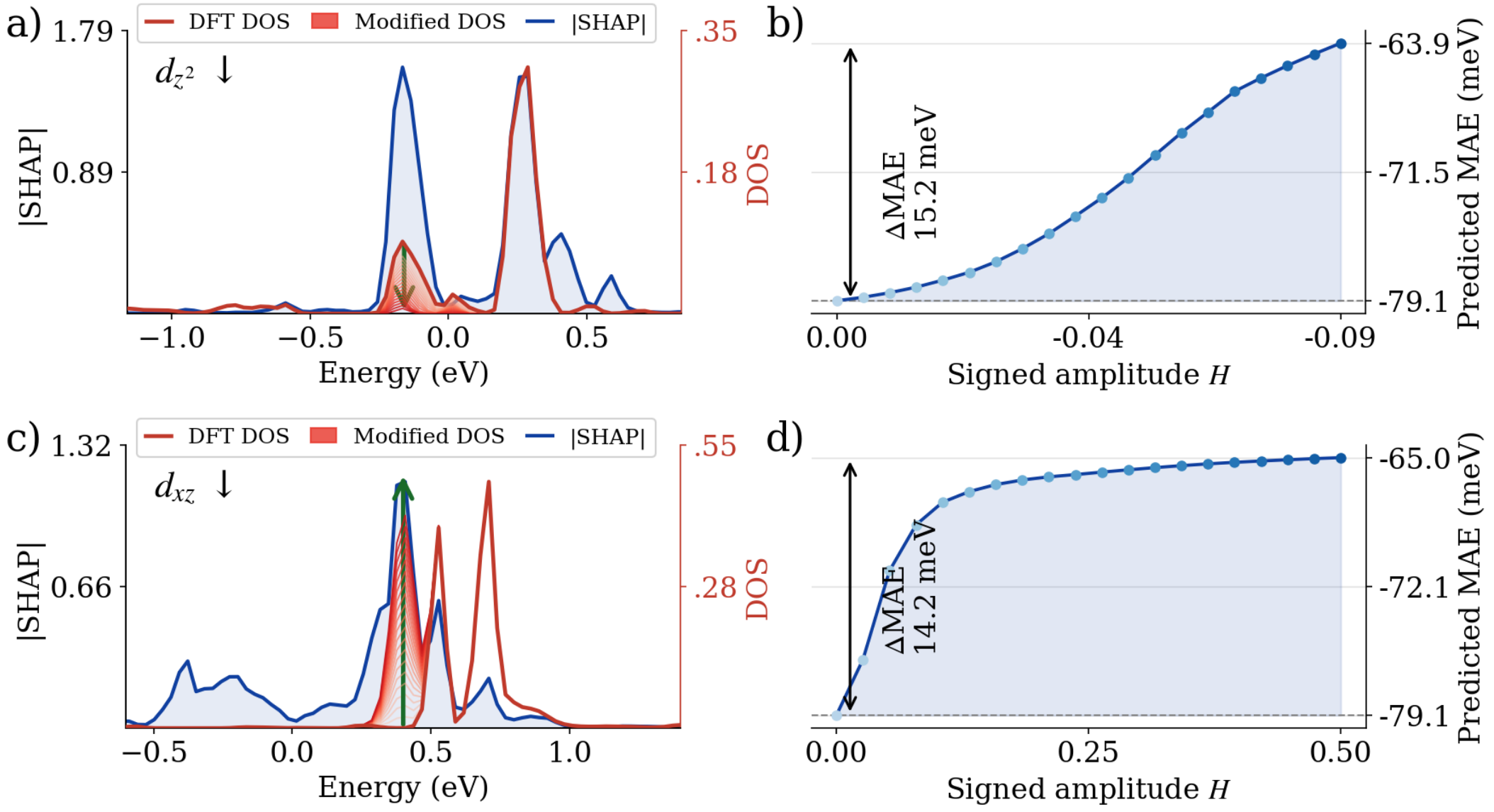


**Figure S19 | Controlled DOS perturbations.** Results are shown for the top TM atom of the strongly negative-MAE OsPt@NSV dimer, with a DFT-computed MAE of −75.7 meV, evaluated using the 1D-CNN. **a**, Attenuation of an occupied DOS feature in a region with strong SHAP attribution; **c**, introduction of spectral weight into another SHAP-active energy window. **b**, **d**, Corresponding MAE predictions as functions of the perturbation amplitude $H$.

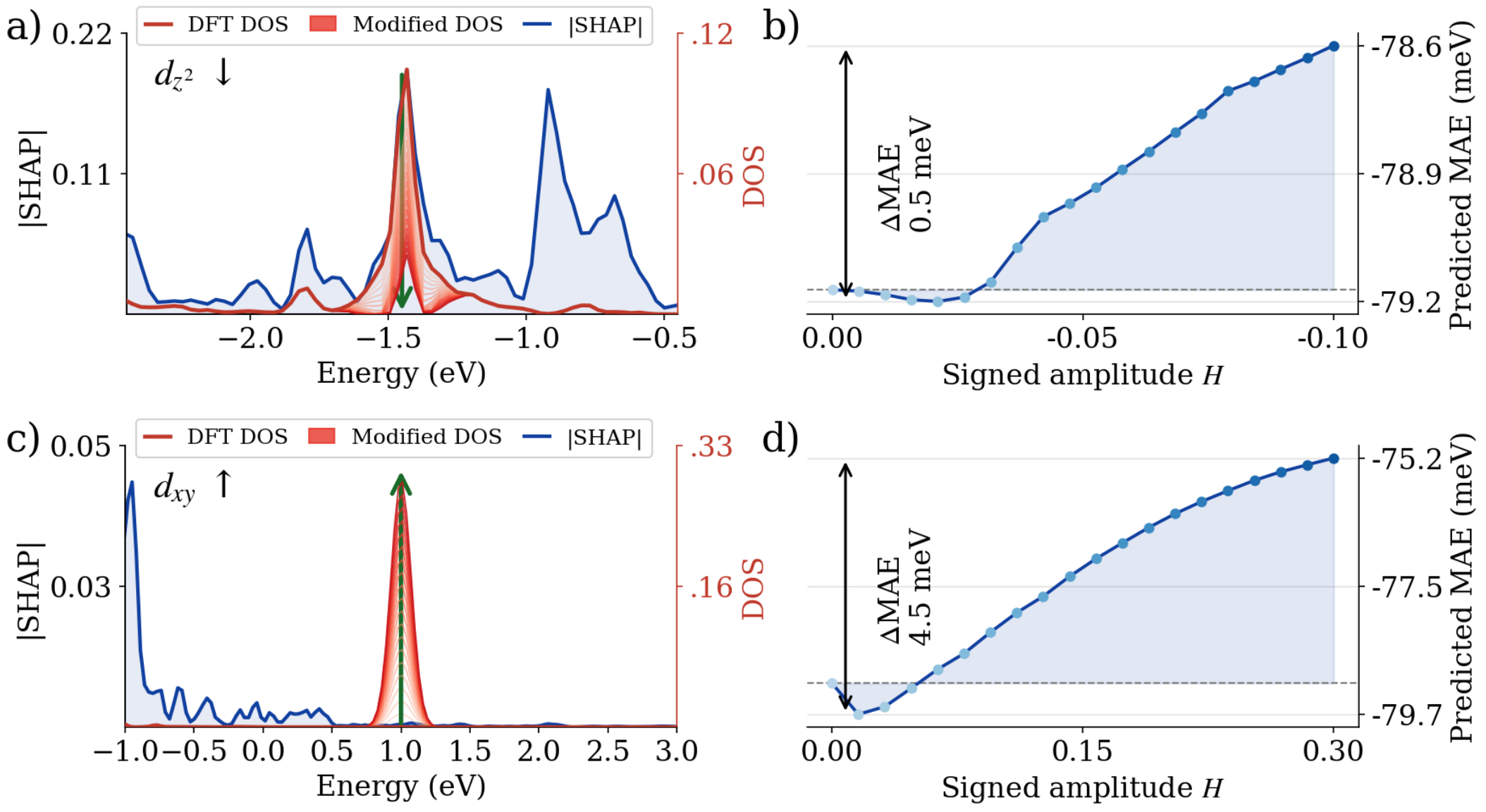


**Figure S20 | Controlled DOS perturbations in SHAP-inactive regions.** Results are shown for the strongly negative-MAE OsPt@NSV dimer, with a DFT-computed MAE of −75.7 meV, evaluated using the 1D-CNN. **a**, Attenuation of a DOS peak for the bottom TM atom; **c**, introduction of an artificial DOS peak for the top TM atom. **b**, **d**, Corresponding MAE predictions as functions of the perturbation amplitude $H$.



\end{document}